\documentclass[fleqn, usenatbib]{mnras}

\usepackage{newtxtext,newtxmath}
\usepackage{subfigure}
\usepackage{graphicx}
\usepackage{bmpsize}
\hypersetup{draft}

\newcommand{\simgt}{\lower.5ex\hbox{$\; \buildrel > \over \sim \;$}}
\newcommand{\simlt}{\lower.5ex\hbox{$\; \buildrel < \over \sim \;$}}

\def\bbeta{\mbox{\boldmath $\beta$}}
\def\btheta{\mbox{\boldmath $\theta$}}

\def\h70kpc{\mathrel{h_{70}^{-1}{\rm kpc}}}

\def\h70Msol{\mathrel{h_{70}^{-1}M_\odot}}

\usepackage[T1]{fontenc}
\usepackage{ae,aecompl}

\usepackage{graphicx}	
\usepackage{amsmath}	
\usepackage{lscape}
\usepackage{comment}
\usepackage{bm}

\title[ShaSS: weak lensing analysis]{Shapley Supercluster Survey:
  mapping the dark matter distribution}

\author[Y.Higuchi et al.]{
Yuchi Higuchi$^{1,2}$\thanks{E-mail: yuichi.higuchi@nao.ac.jp},
Nobuhiro Okabe $^{3,4,5}$,
Paola Merluzzi$^{6}$,
Christopher Paul Haines$^{7}$, \newauthor
Giovanni Busarello$^{6}$, Aniello Grado$^{6}$ and Amata Mercurio$^{6}$
\\
$^{1}$National Astronomical Observatory of Japan, 2-21-1 Osawa, Mitaka, Tokyo 181-8588, Japan\\
$^{2}$Faculty of Science and Engineering, Kindai University, Higashi-Osaka, Osaka, 577-8502, Japan\\
$^{3}$Hiroshima University, 1-3-2 Kagamiyama, Higashi-Hiroshima City, Hiroshima, Japan 739-8511\\
$^{4}$Hiroshima Astrophysical Science Center, Hiroshima University, 1-3-1 Kagamiyama, Higashi-Hiroshima, Hiroshima 739-8526, Japan \\
$^{5}$Core Research for Energetic Universe, Hiroshima University, 1-3-1, Kagamiyama, Higashi-Hiroshima, Hiroshima 739-8526, Japan\\
$^{6}$INAF-Osservatorio Astronomico di Capodimonte, Via Moiariello 16, I-80131 Napoli, Italy\\
$^{7}$Instituto de Astronom\'{i}a y Ciencias Planetarias de Atacama, Universidad de Atacama, Av. Copayapu 485, Copiap\'{o}, Region de Atacama, Chile}
\date{Accepted XXX. Received YYY; in original form 2019}

\pubyear{2019}

\begin{document}
\label{firstpage}
\pagerange{\pageref{firstpage}--\pageref{lastpage}}
\maketitle

\begin{abstract}
We present a 23\,deg$^2$ weak gravitational lensing survey of the
Shapley supercluster core and its surroundings using $gri$ VST images
as part of the Shapley Supercluster Survey (ShaSS). This study reveals the
overall matter distribution over a region containing 11 clusters at
$z{\sim}0.048$ that are all interconnected, as well as several ongoing
cluster-cluster interactions. Galaxy shapes have been measured by using
the Kaiser-Squires-Broadhurst method for the $g$- and $r$-band
images and background galaxies were selected via the $gri$
colour-colour diagram. This technique has allowed us to detect all of
the clusters, either in the $g$-band or $r$-band images, although at
different $\sigma$ levels, indicating that the underlying dark matter
distribution is tightly correlated with the number density of the
member galaxies. The deeper $r$-band images have traced the five
interacting clusters in the supercluster core as a single coherent
structure, confirmed the presence of a filament extending North from
the core, and have revealed a background cluster at $z{\sim}0.17$.  We
have measured the masses of the four richest clusters (A3556, A3558,
A3560 and A3562) in the two-dimensional shear pattern, assuming a
spherical Navarro-Frenk-White (NFW) profile and obtaining a total mass
of $\mathcal{M}_{\rm
  ShaSS,WL}{=}1.56^{+0.81}_{-0.55}{\times}10^{15\,}{\rm
  M}_{\odot}$, which is consistent with dynamical and X-ray studies.
  Our analysis provides further evidence of the ongoing dynamical
evolution in the ShaSS region.
\end{abstract}

\begin{keywords}
gravitational lensing: weak --galaxies: clusters: general --dark matter
\end{keywords}


\section{Introduction}
Large-scale optical and spectroscopic explorations such as the Sloan
Digital Sky Survey \citep{2000AJ....120.1579Y} have revealed the
complex structures of the Universe -- galaxy clusters, filaments and
voids.  Superclusters represent the vastest coherent structures in the
Universe, extending up to $\sim$100\ across as observed in the wide optical surveys \citep[e.g.][]{2011A&A...532A...5E}. 
The observation of superclusters was often considered a challenge to the hierarchical structure formation paradigm since such extremely vast overdense structures, but also the largest voids, are not reproduced by the $N$-body simulations. However, new techniques to analyse $N$-body simulations \citep[e.g.][]{2011MNRAS.413.1311Y, 2019MNRAS.488.5811H} obtained non-zero probabilities of identifying such peculiar systems (overdense and underdense) within redshift surveys.
Recent numerical simulations \citep{2019A&A...623A..97E} showed that the
stability of size and number of superclusters during their evolution
are important properties of the cosmic web and that the number density
of the superclusters thus constrains the cosmological models. 
Also, these structures are still collapsing with galaxy clusters and groups frequently interacting and merging, enhancing the effects of the environment on galaxy evolution. 
These effects are amplified in the supercluster high-density cores, which are gravitationally-bound \citep{2016A&A...595A..70E} and anticipated to become the most massive virialized structures in the distant future.
This implies that galaxy properties and the wider
supercluster evolution are strongly related
\citep[e.g.][]{2009MNRAS.393.1275G, 2009AJ....137.4867L,
  2012ApJ...754..141M, 2016MNRAS.460.3345M, 2018MNRAS.475.4148G,
  2018A&A...620A.149E, 2018MNRAS.478.4336M}.

\begin{figure*}
\centerline{\includegraphics[width=1.5\columnwidth, bb= 0 0 919 606]{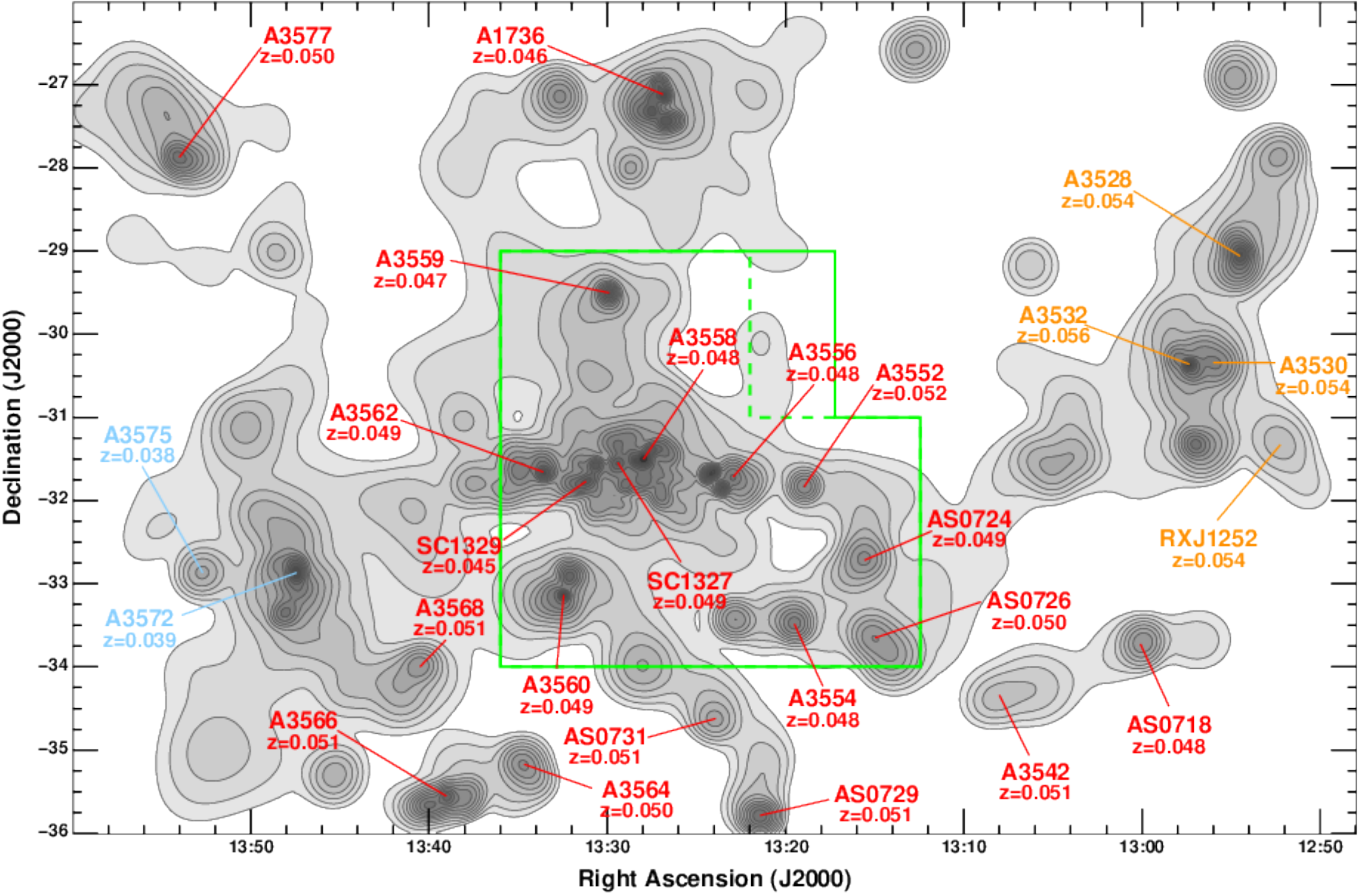}}
\caption{$K$-band luminosity-weighted density map of
  $0.035{\le}z{<}0.060$ galaxies across the full extent of the Shapley
  Supercluster.  $K$-band magnitudes are taken from the 2MASS Extended
  Source Catalog and combines redshifts from the previous wide-field
  spectroscopic surveys covering the region
  \citep[6dFGRS;][]{2009MNRAS.399..683J, 1997A&AS..125..247Q,
    2003MNRAS.339..652K, 2004PASA...21...89D, 2006A&A...447..133P,
    2009A&A...495..707C}.  Abell clusters within the same redshift
  range are labelled, colour coded as: (red) recession velocities
  within 1\,500 km\,s$^{-1}$ of the central cluster Abell 3558; (cyan)
  low-velocity extension $0.035{\le}z{<}0.040$; (orange) high-velocity
  wing, $0.053{\le}z{<}0.060$.  The solid green box outlines the
  23\,deg$^2$ region covered by ShaSS, and the dashed green lines our
  21\,deg$^2$ AAOmega survey (see text). }
\label{fig.2mass}
\end{figure*}

\begin{table*}
\centering
\caption{Properties of the 11 known X-ray galaxy clusters in the
  Shapley supercluster. The central velocities $\langle V_\mathrm{h}\rangle$, velocity dispersions $\sigma_{\nu}$, number of spectroscopic members within $r_{200}$ ($N_\mathrm{z}$), and $r_{200}$ radii (in Mpc) are all determined by \citet{2018MNRAS.481.1055H}. 
  ROSAT-based X-ray bolometric luminosities taken from \citet{2005A&A...444..387D}, except A\,3559, A\,3560 from \citet{1997MNRAS.289..787E}.  
  The last column lists the dynamical masses derived from the analysis of \citet{2018MNRAS.481.1055H}. }
\begin{tabular}{lccr@{$\,\pm\,$}lr@{$\,\pm\,$}lcccr@{$\,\pm\,$}l} \hline
Cluster & R.A.(J2000)& Dec.(J2000)& \multicolumn{2}{c}{$\langle V_\mathrm{h}\rangle$} & \multicolumn{2}{c}{$\sigma_{\nu}$} & $N_\mathrm{z}$ & $r_{200}$ & $L_\mathrm{X,bol}$ & \multicolumn{2}{c}{$\mathcal{M}_\mathrm{dyn}$}\\ 
Name & [degrees] & [degrees] & \multicolumn{2}{c}{[km\,s$^{-1}$]} &
\multicolumn{2}{c}{[km\,s$^{-1}$]} & & [Mpc] &
$10^{43}$[erg\,s$^{-1}]$ & \multicolumn{2}{c}{$[10^{13}{\rm
      M}_\odot]$} \\ \hline AS\,0724 & 198.863788 & -32.710022 &
14\,748 & 74 & 410 & 41 & 29 & 0.945 & 1.5 & 10.06 & 3.02 \\ AS\,0726
& 198.655892 & -33.764558 & 14\,911 & 94 & 603 & 48 & 38 & 1.397 & 0.7
& 32.40 & 7.74 \\ Abell\,3552 & 199.729560 & -31.817560 & 15\,777 & 75
& 334 & 42 & 19 & 0.769 & 0.4 & 5.41 & 2.04 \\ Abell\,3554 &
199.881979 & -33.488139 & 14\,346 & 66 & 602 & 37 & 77 & 1.387 & 2.0 &
31.67 & 5.84 \\ Abell\,3556 & 201.028020 & -31.669858 & 14\,396 & 45 &
628 & 31 & 257 &1.447 & 1.7 & 35.94 & 5.32 \\ Abell\,3558 & 201.986896
& -31.495847 & 14\,500 & 39 & 1007 & 25 & 867 & 2.319 & 66.8 & 148.16
& 13.72 \\ Abell\,3559 & 202.462154 & -29.514386 & 14\,149 & 63 & 521
& 39 & 68 & 1.201 & 1.5 & 20.59 & 4.62 \\ Abell\,3560 & 203.107354 &
-33.135989 & 14\,739 & 52 & 860 & 32 & 304 & 1.981 & 11.7 & 92.29 &
10.30\\ Abell\,3562 & 203.402575 & -31.670383 & 14\,786 & 53 & 769 &
30 & 198 & 1.770 & 33.1 & 65.87 & 7.71 \\ SC\,1327-312 & 202.448625 &
-31.602450 & 14\,794 & 38 & 535 & 17 & 220 & 1.347 & 12.7 & 29.84 &
2.84 \\ SC\,1329-313 & 202.913761 & -31.807260 & 13\,416 & 49 & 373 &
28 & 55 & 0.862 & 5.2 & 7.60 & 1.71 \\ \hline
\end{tabular}
\label{tab.clus_prop}
\end{table*}

The Shapley supercluster (SSC, $z\sim 0.05$) is the largest
conglomeration of Abell clusters in the local Universe (see
Fig.~\ref{fig.2mass}). At its heart there is a complex dense core
consisting of five clusters forming a continuous filamentary structure
2\,degrees ($\sim$8\,Mpc; $H_0=70$\,km\,s$^{-1}$\,Mpc$^{-1}$) in
extent, that is filled with hot gas as seen by both Planck and
$XMM-Newton$ satellites \citep{2014A&A...571A..31P,
  2016MNRAS.460.3345M}.  Across this central region dynamical studies,
X-ray and radio observations showed evidence of multiple
cluster-cluster interactions \citep[e.g.][]{1998MNRAS.300..589B,
  2000MNRAS.312..540B, 1999A&A...341...23K, 2003A&A...402..913V,
  2005A&A...440..867G, 2005AJ....130.2541M}.  Several attempts have
also been undertaken to map the whole supercluster in order to
determine its morphology and mass, as well as ascertain which portions
of the supercluster were gravitationally bound
\citep[e.g.][]{2000AJ....120..523R, 2004PASA...21...89D,
  2005A&A...444..387D, 2006A&A...447..133P, 2008MNRAS.391.1341M}.  The
supercluster resides in the direction of the Cosmic Microwave
Background (CMB) dipole anisotropy.  \citet{1995AJ....110..463Q}
showed that the gravitational pull of the supercluster may account for
$25\%$ of the peculiar velocity of the Local Group required to explain
the anisotropy and their mass would be dominated by inter-cluster dark
matter in that case, while the optical flux distribution lies $\sim25$
degrees away from the CMB dipole.  However, due to a lack of robust
estimates of the SSC mass, the relevance of its gravitational pull
upon the high peculiar velocity (${\sim}$600\,km\,s$^{-1}$) of the
Local Group relative to the Hubble Flow remains an open issue
\citep{1989Natur.342..251R, 2004cgpc.sympE..26K, 2017ApJ...847L...6C}.
An order of magnitude estimate of the SSC mass was provided by
\citet{2000AJ....120..523R}, by means of a dynamical analysis based on
supercluster member galaxies. They estimated the mass of the
central region within $8h^{-1}$Mpc of A\,3558 to be  
$\mathcal{M}\sim10^{16}h^{-1}{\rm M}_\odot$.

Although the previous studies were fundamental to demonstrate the
complex dynamical status of the SSC, the lack of accurate and
homogeneous multi-band imaging and spectroscopic coverage across such
an extended structure prevented, among other things, a quantitative
description of the supercluster environment from filaments to cluster
cores and a robust mass estimate.

With all this in mind we have carried out the Shapley Supercluster
Survey \citep[ShaSS,][]{2015MNRAS.446..803M}, delivering high-quality
optical and near-infrared imaging across a contiguous 23\,deg$^{2}$
(260\,$h^{-2}_{70}$Mpc$^2$) region centred on the supercluster core.
The survey includes nine Abell clusters (A\,3552, A\,3554, A\,3556,
A\,3558, A\,3559, A\,3560, A\,3562, AS\,0724, AS\,0726) and two poor
clusters (SC\,1327-312, SC\,1329-313) whose redshifts all lie within
1\,500\,km\,s$^{-1}$ of Abell\,3558 at $z=0.048$ \citep[see solid
  green box in Fig.~\ref{fig.2mass} and refer for details
  to][]{2015MNRAS.446..803M}. The parameters of the clusters are
given in Table~\ref{tab.clus_prop} \citep[see][]{2018MNRAS.481.1055H}.

The main objective of the ShaSS project is to investigate the role of
cluster-scale mass assembly on the evolution of galaxies, mapping the
effects of the environment in the cluster outskirts and along the
filaments with the aim of identifying the very first interactions
between galaxies and their environment. In order to achieve this goal
it is crucial \textit{to reveal the structure}, i.e. to obtain
detailed maps of the dark matter and baryonic matter distributions
(galaxies, intra-cluster medium), combining galaxy number and stellar
mass density, weak lensing, X-ray and dynamical analyses. In the
present work we characterized the supercluster environment by means of
weak lensing (WL) technique.

While we have been able to produce highly-detailed and complete
two-dimensional density maps of the galaxy distribution across the
Shapley supercluster \citep{2018MNRAS.481.1055H} and the stellar mass
content \citep{2015MNRAS.446..803M}, this stellar content is expected
to only represent a relatively small fraction of the global mass of
this region, in comparison to the hot X-ray emitting gas of the
clusters and the wider dark matter component. More problematically,
the relative fraction of baryonic component that is locked into stars
and galaxies rather than hot gas, and the global mass-to-light ratios
have been shown to vary significantly between galaxy groups and the
most massive clusters, with galaxy groups much more efficient at
converting the baryons into stars and producing light than clusters
\citep[e.g.][]{2005ApJ...618..214T, 2007ApJ...666..147G,
  2013ApJ...778...14G}. Thus it is no trivial matter to translate the
galaxy distribution or stellar mass distribution to the wider mass
distribution.

Weak gravitational lensing enables the overall mass distribution of
clusters and superclusters to be directly measured without requiring
us to make any assumptions regarding the dynamical state of the system
\citep{2003A&A...405..425E, 2004astro.ph.10145O}. The tidal
gravitational field of a cluster leads to the differential deflection of
light coming from background galaxies, distorting them and producing a
coherent shear signal on top of the random intrinsic ellipticities of
individual galaxies. Measuring this coherent distortion pattern among
the background galaxies enables the two-dimensional mass distribution
of the cluster/supercluster to be mapped
\citep[e.g.][]{2010MNRAS.405..257M, 2012MNRAS.420.3213O,
  2014ApJ...795..163U}.  This technique is particularly powerful for
those unvirialised regions beyond the cluster core, where traditional
approaches based on galaxy dynamics or X-ray emission are no longer
suitable.  Weak lensing analyses have been used to detect filamentary
structures connecting adjacent galaxy clusters
\citep{2008MNRAS.385.1431H, 2012Natur.487..202D, 2014MNRAS.441..745H},
or provide mass maps of merging cluster systems
\citep[e.g.][]{2008PASJ...60..345O, 2014ApJ...783...78J}.

In a previous study applying a lensing analysis to superclusters, we
showed the correlation between the early-type galaxy distribution and WL density map in the central $\sim1$\,deg$^2$ region including A\,3558
and SC\,1327-312. We measured for the mass of A\,3558
$\mathcal{M}_{500}=7.63_{-3.40}^{+3.88}\times10^{14}$\,M$_\odot$
consistent with that derived from the X-ray observations
$\mathcal{M}_{500}=(4.62\pm 0.24)\times10^{14}$\,M$_\odot$
\citep{2015MNRAS.446..803M}. The agreement of the two independent
measurements demonstrated the feasibility and effectiveness of the WL
analysis which we will here extend and improve considering the whole
ShaSS region.

Beyond deriving cluster masses using a complementary approach, the WL
technique will enable us to further investigate the nature of the
whole system, tracing the mass distribution outside of the
cluster cores, as well as revealing possible background structures.
In particular, in \citet{2018MNRAS.481.1055H} with the galaxy number
density map and the dynamical analysis we established the
existence of a stream of galaxies connecting A\,3559 to the
supercluster core. Moreover, the updated central redshifts and
velocity dispersions of the 11 clusters confirmed that they all lie
within 1300\,km\,s$^{-1}$ of the central cluster A\,3558. These 11
systems are all inter-connected and lie within a coherent sheet of
galaxies that fills the entire survey region without gaps. Clear
velocity caustics extend right to the survey boundary, indicating that
the entire structure is gravitationally bound and in the process of
collapse. All this invokes and supports a deeper examination. 

The structure of the paper is organized as
follows. Section~\ref{sec:data} describes the main characteristics of
the dataset.  Section~\ref{sec:anamethod} introduces the basics of
weak lensing and the methods of analysis. The shear measurement,
background galaxies selection and fitting procedure are detailed in
Section~\ref{sec:wl_ana}. Section~\ref{sec:result} presents the
results of the WL analysis which are discussed in
Section~\ref{sec:dis}. Lastly, Section~\ref{sec:con} summarizes our
findings.

We adopt the following cosmological parameters: the Hubble parameter
$H_0=70$\,km\,s$^{-1}$\,Mpc$^{-1}$, the density parameter of total matter
$\Omega_{\rm m}=0.3$, $\Omega_\Lambda=0.7$, spectral index $n_s=0.972$
and density fluctuation amplitude $\sigma_8=0.823$, assuming a flat
FLRW cosmology. Otherwise, $h=H_0/100$.

\section{The data}
\label{sec:data}

The ShaSS database consists of high-quality optical $ugri$ imaging
acquired with OmegaCAM \citep{2011Msngr.146....8K} on the 2.6m VLT
Survey Telescope \citep[VST,][]{2011Msngr.146....2C} and near-infrared
$K$-band imaging from the 4.1m Visible and Infrared Survey Telescope
for Astronomy (VISTA), both taking advantage of the exceptional
observing conditions available at Cerro Paranal in Chile. In addition,
our $ad\,hoc$ spectroscopic survey carried out with the AAOmega
spectrograph on the 3.9m Anglo Australian Telescope provides
highly-complete and homogeneous redshift coverage across the full
ShaSS region \citep[see][]{2018MNRAS.481.1055H}.

The corrected OmegaCAM field of view of 1$^\circ$x 1$^\circ$ allows
the whole ShaSS area to be covered with 23 VST fields, sampled at
0.21\,arcsec-per-pixel corresponding to a sub-kiloparsec resolution at
the supercluster redshift. Each of the contiguous ShaSS fields is
observed in four bands, $u$ ($t_\mathrm{exp}$ = 2955\,s), $g$ (1400\,s), $r$
(2664\,s), and $i$ (1000\,s), reaching 5$\sigma$ (AB) magnitude limits
of 24.4, 24.6, 24.1, and 23.3, respectively
\citep[see][]{2015MNRAS.446..803M, 2015MNRAS.453.3685M}. The $r$-band
images were to be acquired in the best seeing conditions with the aim
of using these data for shear measurements and morphological
classification. The median seeing in $r$-band is 0.6\,arcsec,
corresponding to 0.56\,kpc\,$h^{-2}_{70}$ at $z=0.048$.

{\small
\begin{table*}
\begin{center}
\caption{List of the parameters from the $gri$ catalogues that were used to perform the WL analysis.}
  \begin{tabular}{lll}
    \hline
{\bf Parameter} & {\bf Units} & {\bf Description} \\
\hline
$ID$ & & ShaSS identification \\
$RA$ & deg & Right Ascension (J2000) \\
$DEC$ & deg & Declination  (J2000) \\
$MK$ & mag & Kron magnitude \\
$EMK$ & mag & Error on Kron magnitude \\
$MA_{15}$ & mag & Aperture magnitude inside 1.5\,arcsec diameter \\
$EMA_{15}$ & mag & Error on aperture magnitude inside 1.5\,arcsec diameter \\
$MPSF$ & mag & Magnitude resulting from the PSF fitting \\
$EPSF$ & mag & Error on magnitude resulting from the PSF fitting \\   
$SG$ & & Star/galaxy separation \\
$HFF$ & & Halo fraction flag \\
$HF$ & & Halo flag value \\
$SFF$ & & Spike fraction flag \\
$SF$ & & Spike flag value \\
\hline
\end{tabular} 
\end{center}
\label{tab.CATS}
\end{table*}
}

The VST images have been processed and photometrically calibrated
using the VST-Tube pipeline \citep{2012MSAIS..19..362G}, and the catalogue
produced as described in \citet{2015MNRAS.453.3685M}. In each band
the complete catalogues contain a wealth of information.
Table~\ref{tab.CATS} lists the parameters from the catalogues that
have been used for the WL analysis. Both the Kron magnitudes
\citep[$MK$,][]{1980ApJS...43..305K} and the 1.5\,arcsec aperture
magnitudes ($MA_{15}$) were corrected for Galactic extinction
following \citet{2011ApJ...737..103S}.

Since multiple reflections in the internal optics of OmegaCAM can
produce complex image rings and ghosts (hereafter star haloes) near
bright stars, we carefully traced the effects of such features on
source detection \citep[for details see][]{2015MNRAS.453.3685M}.  In
addition, the saturated stars with haloes also affect the images by
producing spike features which are identified. The parameters
$HFF$, $HF$, $SFF$, $SF$ in Table~\ref{tab.CATS} are indicative of the
robustness of the photometry. $HF$ ($SF$) greater than 0 marks that
the source is at least partially affected by the halo (spike) of a
bright star and indicates the star magnitudes (spike strength). While
$HFF$ ($SFF$) \textit{measures} the fraction of the source area
affected by the halo (spike). Finally, the $SG$ parameter identifies
galaxies ($SG=0$) and stars ($SG>0$). 
The saturated stars ($SG=9$) are also indicated.
In the following analysis we excluded all the sources with HF or SF greater than 0, i.e. all the sources even marginally affected by haloes or spikes, as well as saturated stars.

The survey was designed to study the galaxy population down to
$m^\star+6$ at supercluster redshift ($m_r^\star=15$\,mag in AB
magnitudes).  With this precondition, we fixed the required depth in
each band using typical galaxy colours at $z\sim 0.05$ according to
stellar population models \citep[e.g.][]{2003MNRAS.344.1000B}.
Further constraints were placed by (i) the morphological
classification based on the $r$-band imaging which requires a
signal-to-noise ratio about 100\footnote{For the global galaxy
  properties a signal-to-noise ratio about 10-20 is instead
  sufficient.} and (ii) the plan to use $r$-band imaging for the WL
analysis.  This translates in photometric catalogues having different
magnitude limits in the different bands, with the $r$-band catalogue
including the fainter sources in the ShaSS field.

It follows that to maximize the number density of sources for the
background galaxy selection (see Section~\ref{sec:wl_ana}) and the
photometric redshift estimate we take advantage of the deeper $gri$
catalogues. The three catalogues have been cross-correlated using
\textit{STILTS}\footnote{http://www.star.bris.ac.uk/~mbt/stilts/} with
the $r$-band catalogue as reference table and keeping only the sources
detected in all the three bands.

The $g$-band imaging is generally used for the WL mass measurement
presented here, together with the $gri$ catalogues and the
spectroscopic catalogue. Although the $r$-band imaging is deeper, the
$r$-band imaging was found to present significant distortions in some
restricted regions, and thus the $g$-band imaging was used as an
alternative source of shear measurements. For this reason, and to
provide consistency checks across the wider WL maps, the WL analysis
was carried out in both $r$ and $g$ bands.

The
spectroscopic survey was carried out using the AAOmega multi-fiber
spectrograph on the 3.9-m Anglo-Australian telescope, collecting
redshift measurements for 4027 galaxies \citep{2018MNRAS.481.1055H}.
Targets were selected from the VST images as having $i<18.0$ (AB) and
WISE/W$1<15.5$ mag (Vega).  There have also been multiple previous
redshift surveys of the region, and combining these literature
redshifts with our own, results in redshifts being known for 95\% of
all $i<18$, W$1<15.5$ galaxies (5689/6008) across the whole survey
region.  The redshift distribution of the non-SSC galaxies has the
typical form of magnitude-limited samples
\citep[e.g.][]{2009MNRAS.399..683J}, with a median redshift of 0.138
and 95\% of galaxies within the range 0.014--0.295.  This permits
background structures connected to the WL peaks to be reliably
identified and mapped in redshift-space out to $z \sim 0.3$.

\section{Analysis methods}
\label{sec:anamethod}
\subsection{Navarro-Frenk-White model}
Simulations and observational results revealed that the halo density
profile is inversely proportional to the distance $r$ from the cluster
centre inside a scale radius $r_s$, and proportional to $r^{-3}$
outside this radius. This profile is called NFW profile
\citep{1997ApJ...490..493N}, defined by
\begin{equation}
\rho(r)=\frac{\delta_\mathrm{c}\rho_\mathrm{c}}{r/r_\mathrm{s}(1+r/r_\mathrm{s})^2},
\label{eq.nfw}
\end{equation}
where $\rho_\mathrm{c}$ is the critical density of the Universe at the cluster
redshift. The characteristic overdensity is described with the
concentration parameter $c=r_{200}/r_\mathrm{s}$ by
\begin{equation}
\delta_\mathrm{c}=\frac{200}{3}\frac{c^3}{{\rm ln}\left(1+c\right)-c/\left(1+c\right)}, 
\end{equation}
where $r_{200}$ is defined as the radius inside which the mass density of
the halo is equal to 200 times the critical density. The scale radius
can be derived from the mass of the halo $M_{200}$ and the
concentration parameter as
\begin{equation}
r_s = \left( \frac{3M_{200}}{800\pi\rho_\mathrm{c}} \right)^{1/3}c.    
\end{equation}
Therefore, the density profile can be determined as a function of mass and
concentration parameter when the halo redshift is known.

\subsection{Concentration and mass relation}
$N$-body simulations and observational results show that halo mass
is correlated with the concentration parameter
\citep{2000ApJ...535...30J, 2001MNRAS.321..559B, 2009ApJ...707..354Z,
  2012MNRAS.420.3213O, 2014ApJ...795..163U}.  The concentration and
mass relation (hereafter, \textit{c-M} relation) can be modeled with
four parameters $M_{\rm piv}$, $A$, $B$ and $C$ at a redshift of $z$
\citep{2008MNRAS.390L..64D} as
\begin{equation}
    c = A\left(\frac{M_{200}}{M_{\rm piv}}\right)^B(1+z)^C.
    \label{eq.cm-relation}
\end{equation}
In the following analysis, we used $M_{\rm piv}=2\times10^{12}h^{-1}{\rm M}_\odot$, $A=7.85$, $B=-0.081$ and $C=-0.71$ which were obtained by fitting haloes in the $N$-body simulation \citep{2008MNRAS.390L..64D}.

\subsection{Lensing equations}
The three-dimensional potential $\Phi(D_\mathrm{d}{\bf \theta}, z)$ of a lens is related to the lensing potential $\psi({\bf \theta})$ on the sky plane as
\begin{equation}
\psi(\bm{\theta}) = \frac{2D_\mathrm{ds}}{c^2 D_\mathrm{d} D_\mathrm{s}}\int dz \Phi(D_\mathrm{d}\bm{\theta}, z),
\end{equation}
where $\bm{\theta}$ is a position vector on the sky plane and
$c$ is the speed of light.
$D_\mathrm{d}$, $D_\mathrm{s}$ and $D_\mathrm{ds}$ are respectively the angular diameter
distances between the observer and the lens, the observer and the
sources, and between the lens and the sources.
Given the cluster properties, the convergence and shear for the NFW profile can be obtained as the derivatives of the lens potential.
The potential is related to the shear as
\begin{equation}
\gamma_1(\bm{\theta})=\frac{1}{2}\left(\frac{\partial^2\psi(\bm{\theta})}{\partial\theta_1^2}-\frac{\partial^2\psi(\bm{\theta})}{\partial\theta_2^2} \right),
\end{equation}

\begin{equation}
\gamma_2(\bm{\theta})=\frac{\partial^2\psi(\bm{\theta})}{\partial\theta_1\partial\theta_2}.
\end{equation}
For spherically-symmetric objects, the direction of shear has only a tangential component.
It is useful to define the tangential $\gamma_+$ and cross $\gamma_\textrm{x}$ components for $\gamma_1$ and $\gamma_2$ as 
\begin{equation}
\left( \begin{array}{c}
     \gamma_+\\
     \gamma_\textrm{x}
\end{array}
\right)
=
\left( \begin{array}{cc}
\textrm{-cos}2\phi & \textrm{-sin}2\phi  \\ 
\textrm{-sin}2\phi & \textrm{cos}2\phi
\end{array} \right) \left( \begin{array}{c}
\gamma_1 \\ 
\gamma_2
\end{array} \right), 
\label{eq.defshear}
\end{equation}
where $\phi$ is the angle between the galaxy position and $\theta_1$ axis. 
The shear profile for the spherical symmetric NFW profile can be written  as a function of the distance from the centre by \citep{2000ApJ...534...34W, 2001PhR...340..291B}
\begin{eqnarray}
          && \frac{r_\mathrm{s}\delta_\mathrm{c}\rho_\mathrm{c}}{\Sigma_\mathrm{c}}\left[ \frac{8{\rm arctanh}\sqrt{(1+x)/(1-x)}}{x^2\sqrt{1-x^2}} + \frac{4}{x^2}{\rm ln\left(\frac{x}{2}\right)}\right.\nonumber\\
          &&\left.-\frac{2}{x^2-1}+\frac{4{\rm arctanh}\sqrt{(1-x)/(1+x)}}{(x^2-1)\sqrt{1-x^2}}\right]~~(x<1),\nonumber\\
\gamma_+(x) &=& \frac{r_\mathrm{s}\delta_\mathrm{c}\rho_\mathrm{c}}{\Sigma_\mathrm{c}}\left[\frac{10}{3}+4{\rm ln}\left(\frac{1}{2}\right)\right]~~~~~~~~~~~~~~~~~~~(x=1),\\
          &&\frac{r_\mathrm{s}\delta_\mathrm{c}\rho_\mathrm{c}}{\Sigma_\mathrm{c}}\left[ \frac{8{\rm arctanh}\sqrt{(x-1)/(1+x)}}{x^2\sqrt{x^2-1}} + \frac{4}{x^2}{\rm ln\left(\frac{x}{2}\right)}\right.\nonumber\\
          &&\left.-\frac{2}{x^2-1}+\frac{4{\rm arctanh}\sqrt{(x-1)/(1+x)}}{(x^2-1)^{3/2}} \right] ~~(x<1)\nonumber,
\label{eq.nfwshear}
\end{eqnarray}
where $x$ is a scaled radius defined by $x=R/r_\mathrm{s}$. 
$R$ is a distance on the sky plane, defined as $R=D_\mathrm{d}\sqrt{\theta_1^2+\theta_2^2}$.
From observations, we can only obtain the reduced shear defined by
\begin{equation}
    g_\alpha=\frac{\gamma_\alpha}{1-\kappa},
\end{equation}
where $\alpha$ takes the values of 1 or 2 for each shear component and
$\kappa$ is the convergence.

In this study, we measured cluster properties in the two-dimensional plane following \citet{2010MNRAS.405.2215O} and \citet{2010PASJ...62..811O}.
We divided the sky into pixels. 
Then we calculated the average shear value at each pixel, which is obtained at a position $\bm{\theta}$ by 
\begin{equation}
\left<g_\alpha\right>\left(\bm{\theta}\right) = \frac{\sum_{i=1}^N w_i g_{\alpha,i}\left(\bm{\theta}_i\right)}{\sum_{i=1}^{N} w_i},
\label{eq.aveg}
\end{equation}
where $\bm{\theta}_i$ is the position of the $i$-th source galaxy in the pixel.
The weight for the $i$-th galaxy is defined by
\begin{equation}
w_i=\frac{1}{\sigma_{\mathrm{e},i}^2+\Delta^2}.\label{eq:weight}
\end{equation}
$\sigma_{\mathrm{e}, i}$ is a shape measurement error for the $i$-th galaxy.
We used $\Delta=0.4$ through this paper. 
The shape measurement error for a pixel is obtained as
\begin{equation}
    \sigma_\mathrm{e}^2=\frac{\sum_{i=1}^N \left(w_i \sigma_{\mathrm{e}, i}\right)^2 }{\left(\sum_{i=1}^N w_i\right)^2}.
    \label{eq.sigmae}
\end{equation}

\subsection{Covariance matrix}
In order to fit the cluster profiles with the NFW profile, we calculated the chi-square value for each parameter estimation step, defined as
\begin{eqnarray}
    \chi^2 = \sum_{\alpha, \beta=1}^2\sum_{k, l=1}^{N_\textrm{pixel}} \left[g_\alpha\left(\bm{\theta}_k\right) - g_{\alpha}^\textrm{model}\left(\bm{\theta}_k\right) \right]\bm{C}^{-1}_{\alpha\beta, kl}\nonumber \\
    \times\left[ g_{\beta}\left(\bm{\theta}_l\right) - g_{\beta}^\textrm{model}\left(\bm{\theta}_l\right) \right],
\end{eqnarray}
where $g^\textrm{model}\left(\bm{\theta}_k\right)$ is the model value for a given parameter set at a position $\bm{\theta}_k$ and $N_\textrm{pixel}$ is the number of pixels.
$\bm{C}^{-1}$ is the inverse covariance matrix.
In the estimation of the covariance matrix, we took into account the contribution from the shape noise $\bm{C}^\textrm{shape}$ and large-scale structures $\bm{C}^\textrm{LSS}$:
\begin{equation}
    \bm{C} = \bm{C}^\textrm{shape} + \bm{C}^\textrm{LSS}.
\end{equation}
Assuming that the ellipticities of galaxies are not correlated, the covariance matrix term for the shape noise is estimated with equation~(\ref{eq.sigmae}) as
\begin{equation}
    \bm{C}_{\alpha\beta, kl}^\textrm{shape} = \sigma_e^2\left(\bm{\theta}_k\right)\delta^K_{\alpha\beta}\delta^K_{kl},
\end{equation}
where $\delta^K$ is the Kronecker delta function.
The covariance matrix for large-scale structures can be estimated as
\begin{equation}
    \bm{C}^\textrm{LSS}_{\alpha\beta, kl} = \xi_{\alpha\beta}\left(\theta\right),
\end{equation}
where $\xi\left(\theta\right)$ is the cosmic shear correlation function.
We assumed the correlation function depends only on the distance between the galaxy positions.
Each component of the shear correlation function is calculated as \citep{2001PhR...340..291B, 2011ApJ...738...41U, 2016ApJ...821..116U}
\begin{eqnarray}
\xi_{11}\left(\theta\right) & = & \frac{1}{2}\int \frac{ldl}{2\pi}P_\kappa\left(l\right)\left[J_0\left(l\theta\right)+J_4\left(l\theta\right)\textrm{cos}\left(4\phi\right)\right],\nonumber\\
\xi_{22}\left(\theta\right) & = & \frac{1}{2}\int \frac{ldl}{2\pi}P_\kappa\left(l\right)\left[J_0\left(l\theta\right)-J_4\left(l\theta\right)\textrm{cos}\left(4\phi\right)\right],\\
\xi_{12}\left(\theta\right) & = & \frac{1}{2}\int \frac{ldl}{2\pi}P_\kappa\left(l\right)J_4\left(l\theta\right)\textrm{sin}\left(4\phi\right),\nonumber
\end{eqnarray}
where $J_{0, 4}$ are respectively the zeroth and fourth order Bessel
functions, and $P_\kappa\left(l\right)$ is the convergence power spectrum.

\section{Shape measurement and source galaxy selection}
\label{sec:wl_ana}


The shapes of galaxies were measured using the KSB method
\citep{1995ApJ...449..460K}, with some modifications
\citep[see][]{2010ApJ...714.1470U, 2012MNRAS.420.3213O,
  2013ApJ...769L..35O, 2014ApJ...784...90O, 2016MNRAS.456.4475O} in
both $g$ and $r-$band imaging. Image ellipticity was derived from the
weighted quadruple moments of the surface brightness of objects. The
data region of each pointing was divided into several rectangular
blocks based on the typical coherent scale of the measured PSF
anisotropy pattern \cite[e.g.][]{2016MNRAS.456.4475O}. 
We selected bright unsaturated stars in the half-light radius, $r_\mathrm{h}$--magnitude plane to estimate the stellar anisotropy kernel,
$q^*_{\alpha} = (P^*_{{\rm sm}})^{-1}_{\alpha \beta}e_*^{\beta}$.
$P_{\rm sm}^{\alpha\beta}$ is the {smear polarisability} matrix.
$e_{\alpha}$ is the image ellipticity. Quantities with an asterisk
denote those for stellar objects. We corrected the PSF anisotropy
with the equation
\begin{equation} 
e'_{\alpha} = e_{\alpha} - P_{\rm sm}^{\alpha \beta} q^*_{\beta}. 
\label{eq:qstar}
\end{equation}

\begin{figure}
    \centering
    \includegraphics[width=1.0\columnwidth]{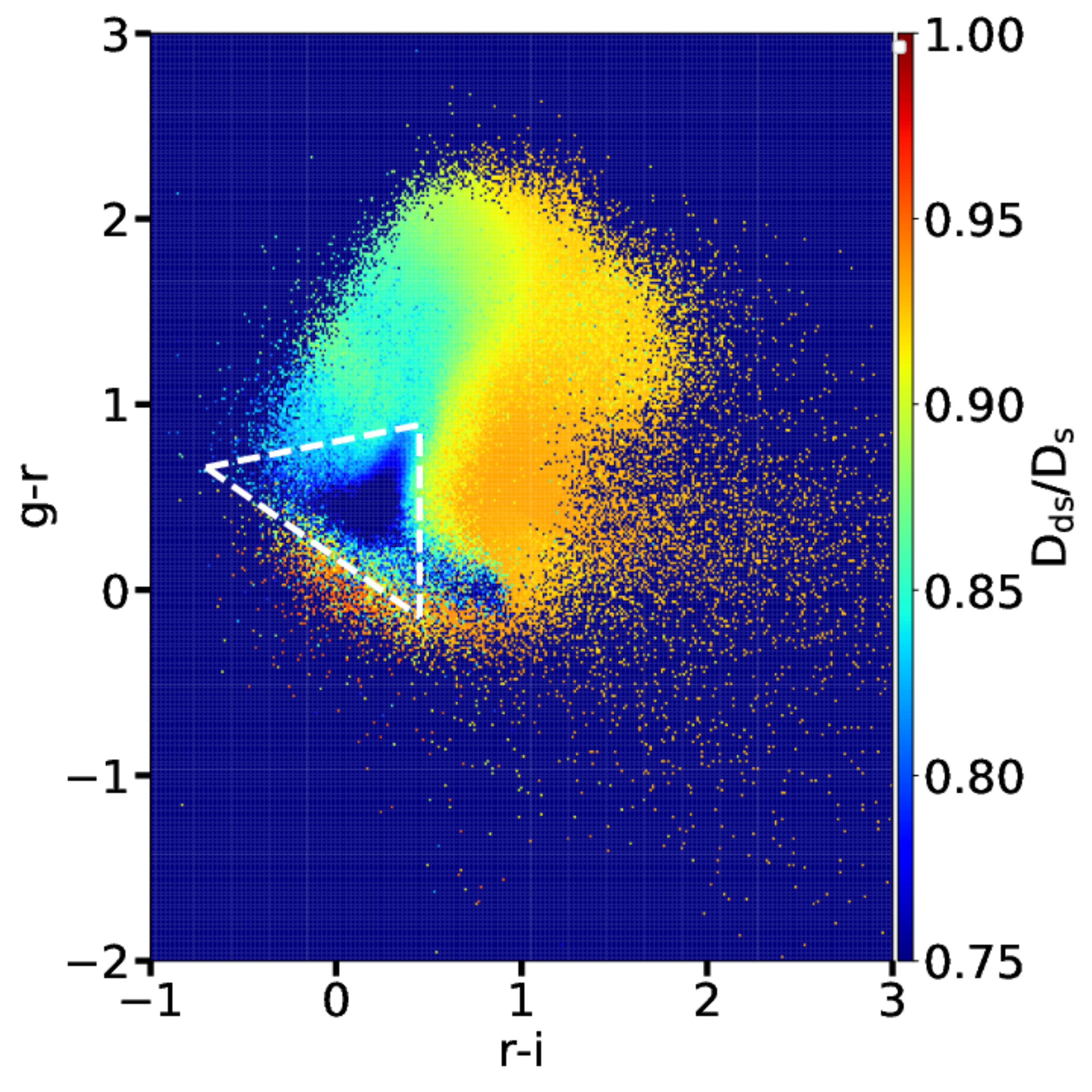}
    \caption{Ratio $D_\mathrm{ds}/D_\mathrm{s}$ of the distances between the observer
      and the sources and the lens and the sources in the
      colour-colour diagram. The horizontal axis shows \textit{r-i}.
      The vertical axis shows \textit{g-r}. Colour indicates the mean
      value of $D_\mathrm{ds}/D_\mathrm{s}$ in each pixel.  The galaxies inside the
      triangle region are defined as the foreground and cluster member
      galaxies.}
    \label{fig.ccmap}
\end{figure}

\begin{figure}
    \centering
    \includegraphics[width=1.0\columnwidth]{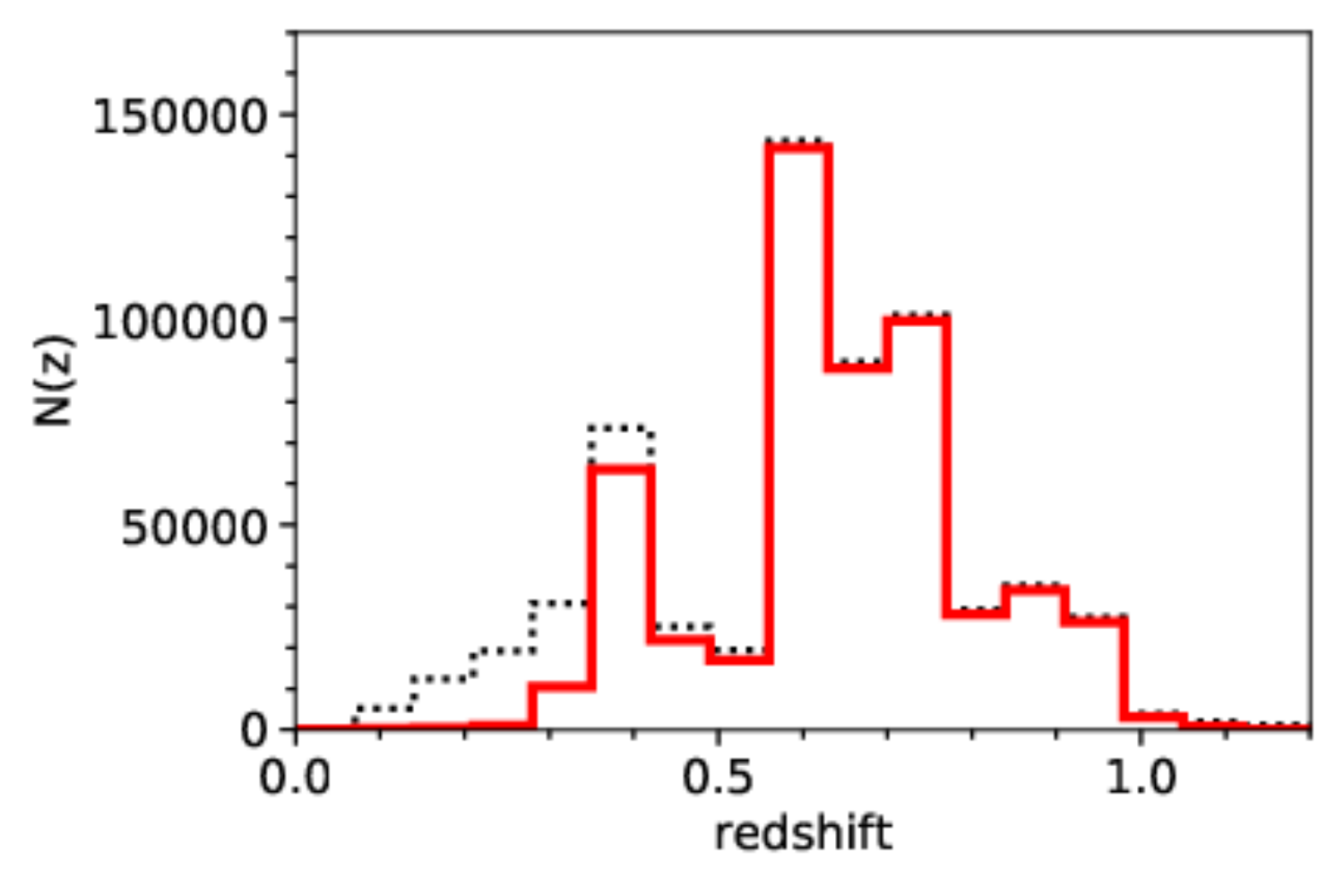}
    \caption{
    Redshift distribution of the source galaxies.
    The horizontal axis shows the redshift.
    The vertical axis shows the number count.
    We only used galaxies whose $i$-band magnitudes are between 21.5 and 24.5.
    The dotted line shows its distribution before adopting the selection criteria for the background galaxy.
    The solid line shows the result after adopting the criteria.
    The weighted mean redshift is 0.81.
    }
    \label{fig:source_dis}
\end{figure}

We estimated $q^*_{\alpha}(\btheta)$ at each galaxy position,
$\btheta$, by using as a fitting function second-order bi-polynomials
of the vector $\btheta$ with iterative $\sigma$-clipping rejection.
Since the PSF distortion pattern in the VST data is locally variable,
we carried out star and galaxy separation in each rectangular block.
We then calibrated the KSB isotropic correction factor for individual
objects using a subset of galaxies detected with high significance
$\nu>50$. The isotropic PSF calibration was also carried out in the
individual blocks used for the anisotropic PSF correction. We checked
the ellipticities for galaxies detected over different pointings and
found a general agreements. We then adopted the average of the
ellipticity for the weak lensing mass measurement with the weight of
equation~(\ref{eq:weight}).

The redshift of each galaxy in the $gri$ ShaSS photometric catalogues was estimated by using the COSMOS photometric redshift catalogue
\citep{2013A&A...556A..55I}. We first computed the lensing kernel,
$\beta_i\equiv{D_{\mathrm{ds},i}}/{D_{\mathrm{s},i}}$, of the $i-$th galaxy as an
ensemble average of the $N$ nearest neighbours in the $gri-$ magnitude
space of the $j$-th galaxy in the COSMOS catalogue,
\begin{eqnarray}
  \beta_i=\langle{D_\mathrm{ds}}/{D_\mathrm{s}}\rangle_{\rm COSMOS}=\frac{1}{N}\sum_j^N{D_{\mathrm{ds},j}}({z_\mathrm{s}})/{D_{\mathrm{s},j}}({z_\mathrm{s}}). \label{eq:beta}
\end{eqnarray}
We then assigned the redshift of each galaxy from $\beta_i$, where we adopted $N=50$. 

To define the foreground and background galaxies, we used the distance
ratio in the colour-colour diagram. Fig.~\ref{fig.ccmap} shows the
mean distance ratio in the colour-colour diagram.  In the map, we
divided the diagram into cells and calculated the mean value in each
pixel.  As seen in Fig.~\ref{fig.ccmap}, there is a clump having the
low value which corresponds to the cluster members and foreground
galaxies. In order to exclude such galaxies, we defined the region
residing in the clump as
\begin{eqnarray}
g-r&<& 0.2(r-i)+0.8 ~({\rm for}~-0.4\leq r-i \leq0.3), \\
g-r&>& -0.7(r-i)+0.17 ~({\rm for}~-0.4\leq r-i \leq0.3),\\
r-i&<& 0.45 ~({\rm for}~0.17\leq g-r \leq 0.63).
\end{eqnarray}
To exclude the effects from bright stars on the shape measurement of
the galaxies, we did not use galaxies which were closer than one star half-light radius from the brightest (<18 mag) stars.
Moreover, we only used the galaxies between
$r=21.5$ and $r=24.5$ magnitude. We could not find large dependence of
shear profiles on the selection criteria and the magnitude of the
bright stars. After adopting these cuts, the number density of
galaxies for the shear measurement becomes $n_g=7$ arcmin$^{-2}$.
Fig.~\ref{fig:source_dis} shows the redshift distribution for galaxies between $i=21.5-24.5$ magnitude before (dotted line) and after adopting the selection criteria (solid line).  
We can see that galaxies at lower redshifts are eliminated by adopting the colour cut. 
As a result, the mean source redshift weighted by
the estimated errors of $D_{\rm ds}/D_{\rm s}$ becomes $z_s=0.81$ after the colour cut.

\section{Results}
\label{sec:result}
\subsection{Weak lensing mass reconstruction}

\begin{figure*}
\includegraphics[width=120mm]{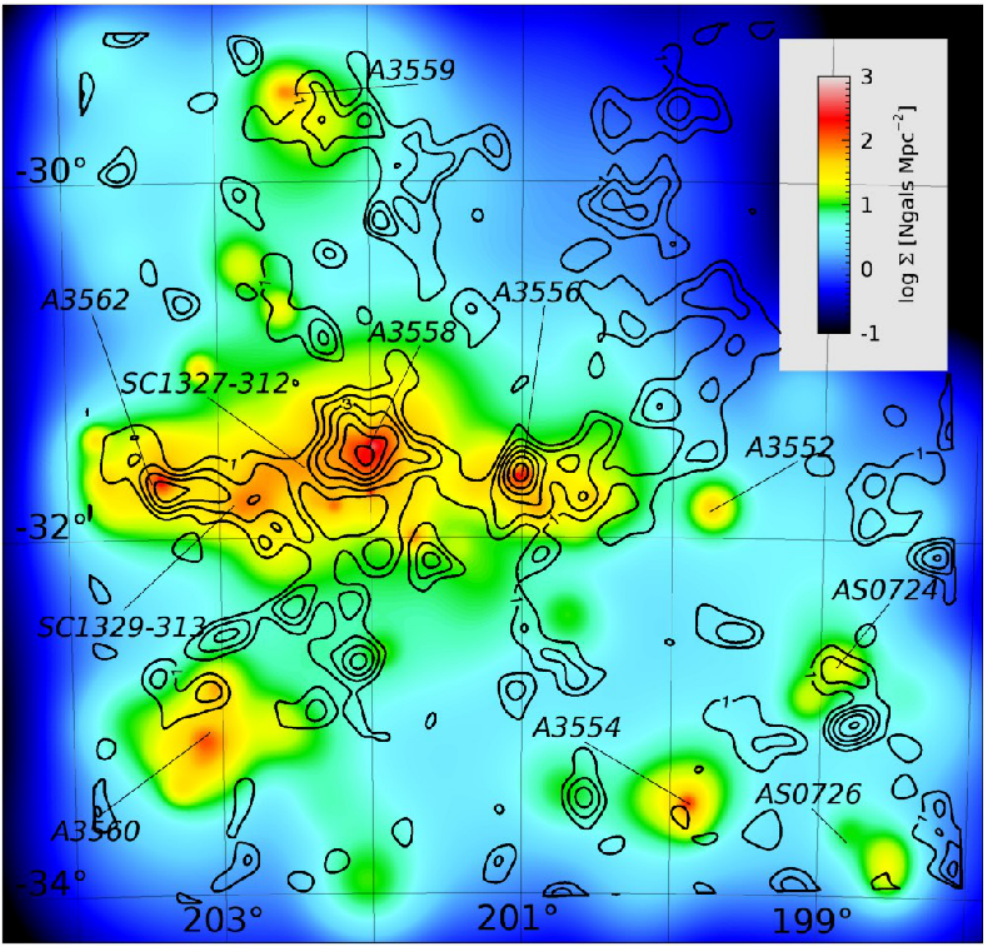}
\includegraphics[width=120mm]{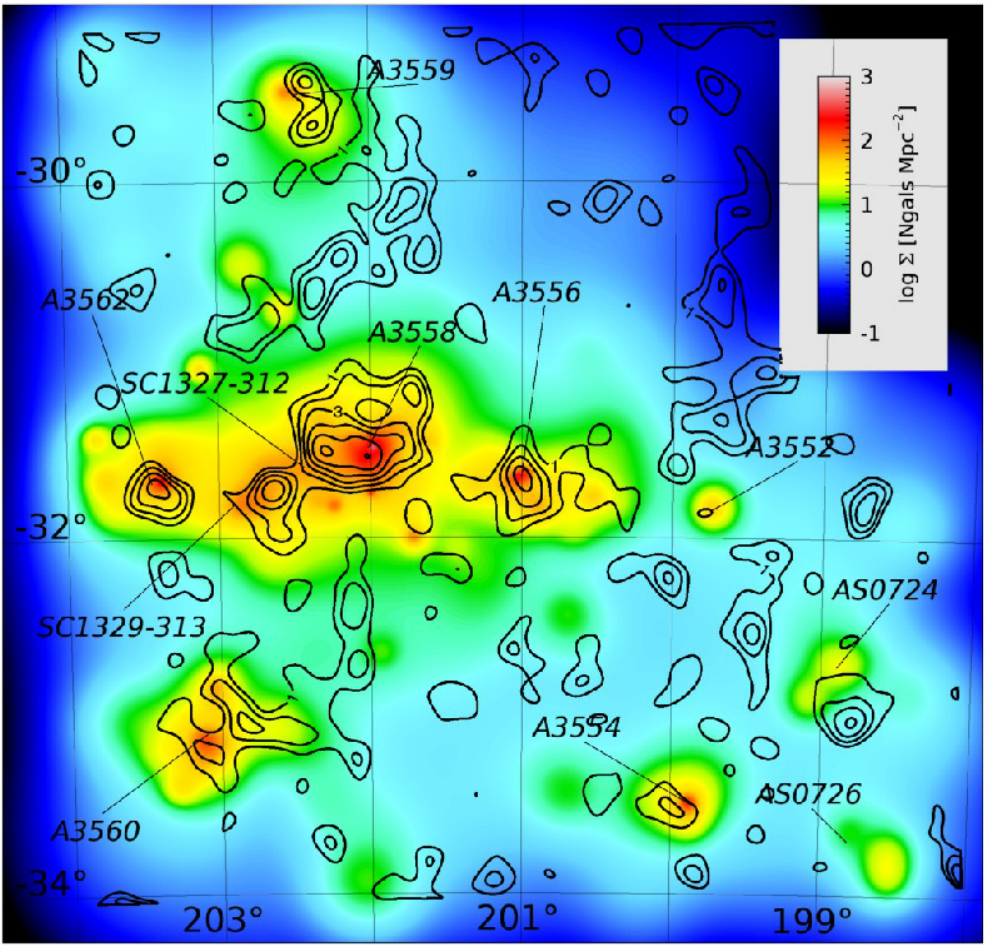}
\caption{Galaxy number density from \citet{2018MNRAS.481.1055H} colour-coded as
  indicated in the right bar and with overlaid the contours of WL mass
  map as derived using the $r$-band imaging (upper panel) and $g$-band
  imaging (lower panel).
  The overlaid contours are stepped by $1\sigma$ from $1\sigma$.}
\label{fig.WL_dens}
\end{figure*}

We reconstructed the projected mass distribution, the so-called mass
map, by following \cite{1993ApJ...404..441K, 2008PASJ...60..345O}. We
employed a Gaussian smoothing scale of FWHM$=10$ arcmin.
Fig.~\ref{fig.WL_dens} shows the WL mass map (contours) of the Shapely
supercluster region for $r$ and $g-$bands, respectively. The errors
were estimated by a bootstrap realization ($N=10000$) which randomly
rotates galaxy orientations with fixed positions. The colours in the
figures showed the galaxy number density derived from
\citet{2018MNRAS.481.1055H}. The WL mass distribution is highly
associated with the spatial distribution of Shapley supercluster
member galaxies (see Fig.~\ref{fig.WL_dens}). In particular, we find
significant peaks around A\,3556, A\,3558, SC\,1327-312, SC\,1329-313,
A\,3562, and A\,3559 in both $r$- and $g$-band images. Less pronounced
peaks are associated with A\,3560 and A\,3554 in the $g$-band images,
where a marginal detection of A\,3552 is also present. The lack of WL
signal in $r$-band images for the cluster A\,3560 can be explained
with the peculiar distortion affecting the $r$-band images of this VST
field. On the other hand AS\,0724 seems to be detected only in the
$r$-band images. We could not find instead a clear peak around
AS\,0726 neither in the $r$- nor the $g$-band images.

Some of the mass peaks without any adjacent overdensity in
supercluster galaxies are associated with background components (see
Section~\ref{sec:dis}). Therefore, our mass map is reasonably
explained by the superposition of the supercluster component and
background structures.

\begin{figure*}
    \centering
    \includegraphics[width=2\columnwidth]{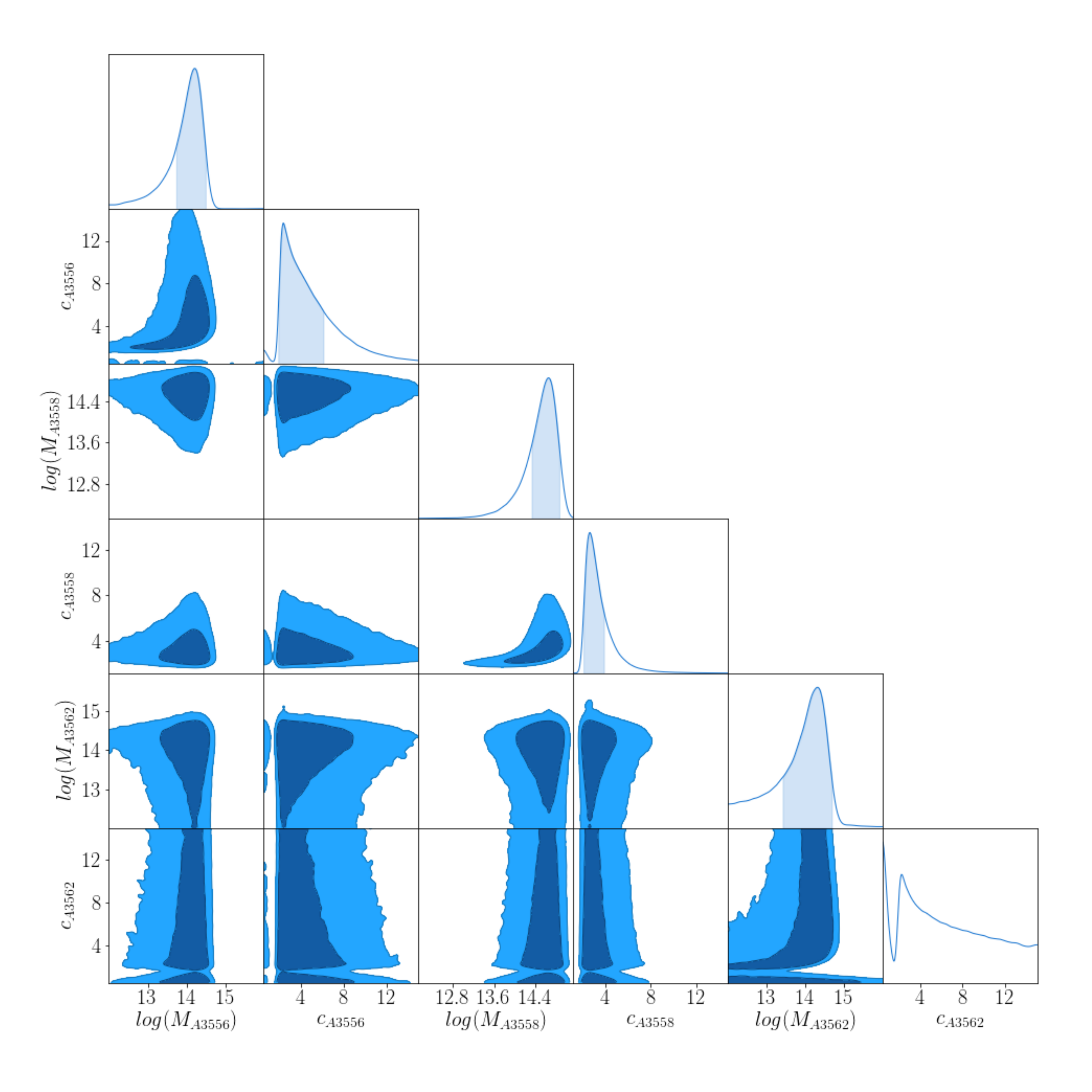}
    \caption{
    MCMC results for A\,3556, A\,3558 and A\,3562 with a pixel scale of 3 arcmin.
  The horizontal axis shows mass $M_{200}$ in units of [$h^{-1}\mathrm{M}_\odot$].
  The vertical axis shows concentration parameter.
  Contours are stepped by $1\sigma$.
    }
    \label{fig:mcmc_3arc_3clus}
\end{figure*}

\begin{figure}
    \centering
    \includegraphics[width=1\columnwidth]{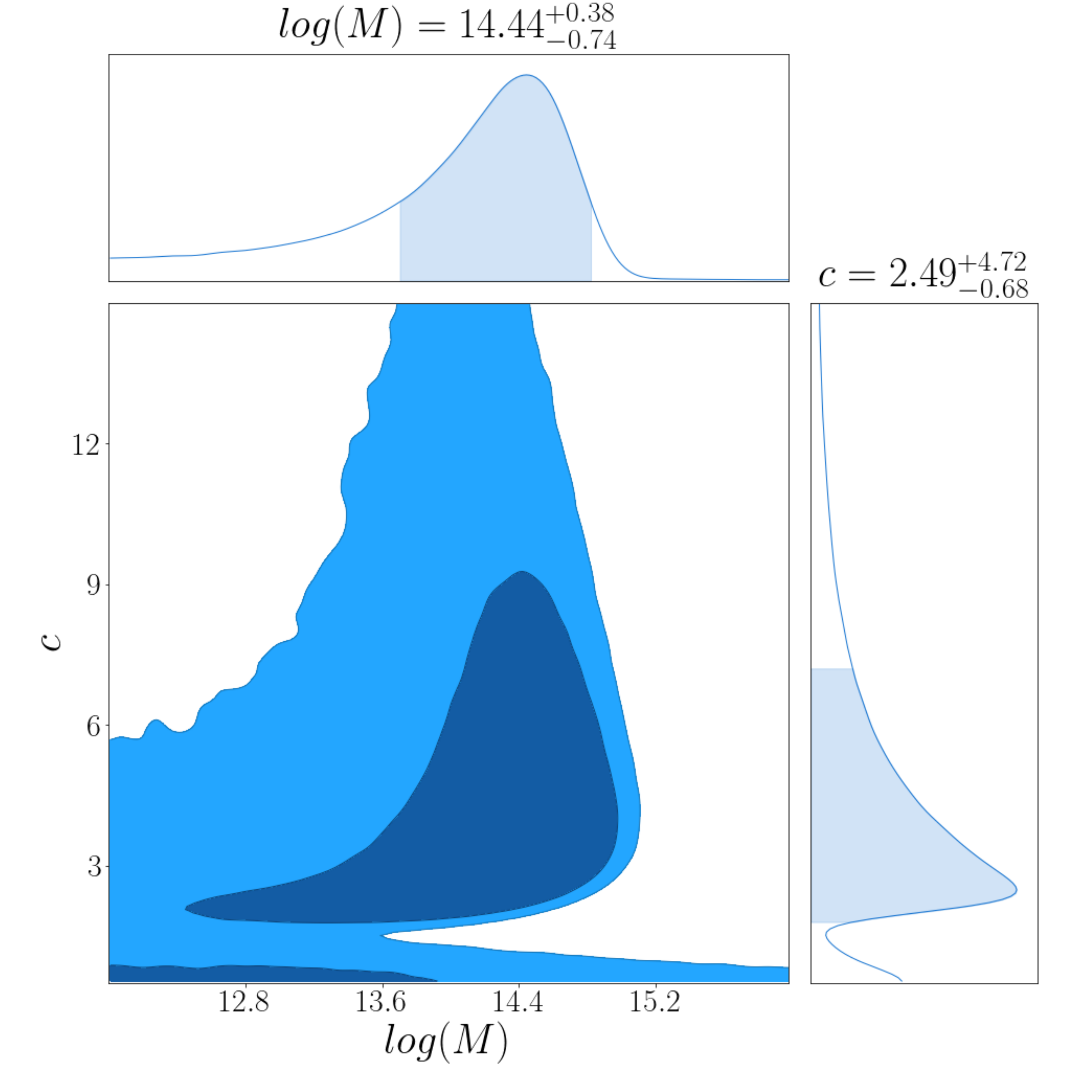}
    \caption{
    MCMC results for A\,3560 with a pixel scale of 3 arcmin.
  The horizontal axis shows mass $M_{200}$ in units of [$h^{-1}\mathrm{M}_\odot$].
  The vertical axis shows concentration parameter.
  Contours are stepped by $1\sigma$.
    }
    \label{fig:mcmc_3arc_fix}
\end{figure}


\subsection{Fitting with the NFW profile}
To estimate the mass and concentration parameters for each cluster, we
fitted the shear values with the NFW profile on a two dimensional
shear map.  Due to the shape measurement problem, we could not derive
the properties of the individual low mass clusters from the fitting.
Therefore, the two-dimensional fitting was carried out only for the
most massive clusters A\,3556, A\,3558, A\,3560 and A\,3562 (see
Table~\ref{tab.clus_prop}). 
We simultaneously fitted the three clusters A\,3556, A\,3558 and A\,3562 by assuming three NFW profiles.
The cluster A\,3560, which was far from the three clusters, was also fitted in the two-dimensional plane independently.  
On the other hand, we measured the
tangential shear profiles for each low-mass cluster, and derived their
mass and concentration parameters. Moreover, we stacked the shear
profiles to investigate the average properties of the low-mass
clusters.

In the two-dimensional fitting, we selected galaxies in the range of
$200.3\leq \textrm{R.A.} \leq 203.8$ and $-32.1 \leq \textrm{Dec.}
\leq -31.1$ for the analysis of A\,3556, A\,3558 and A\,3562.  
For A\,3560, galaxies in the range of $202.6\leq \textrm{R.A.}\leq203.6$
and $-33.6\leq\textrm{Dec.}\leq-32.6$ were instead used. 
The selected regions were divided into cells with the pixel scale of $3$ arcmin on a side.  
We calculated a shear value in each pixel by following equation~(\ref{eq.aveg}). 
In the fitting, we used the positions of the brightest cluster galaxies (BCG) as the centres of the clusters.
We ran the Markov chain Monte Carlo (MCMC) by parametrizing masses and
concentration parameters.  
For analysis of the three clusters (A\,3556, A\,3558 and A\,3562), we searched the best fitted parameters for each clusters simultaneously.  
In order to get the best fitted parameters from the chain distribution, we used the software \textsf{ChainConsumer} \citep{Hinton2016}.
We also carried out the fitting for the cluster A\,3560 independently.

In the calculation of the covariance matrix, we assumed that source
galaxies are present at the average source redshift of the background
galaxies. Since the ShaSS area is large, we only used the diagonal
terms of the covariance matrix to reduce the computational time in the
fitting. We searched the best fit parameters in the range
$10^{13}M_\odot\leq M_{200} \leq10^{16}M_\odot$ and $0.5 \leq c \leq
15$. In order to check the effect of the pixel scale, we also ran the
MCMC for the pixel scale of 2 arcmin and 4 arcmin, respectively. The
obtained results were consistent with those for 3\,arcmin within
$2\sigma$.
Moreover, we derived the masses for the $r-$band data.
However, the obtained results are consistent with the masses estimated by $g$-band data within $2\sigma$. 
In addition, we also added the centres of the clusters as parameters for the fitting.
We searched the centres within 5 armin from the positions from the BCGs.
However, the offset between obtained centres and the positions of the BCGs were within a few arcmin while the centre for A\,3560 did not converge.
The differences for the other parameters were consistent within the error bar even in this case. 
Fig.~\ref{fig:mcmc_3arc_3clus} and \ref{fig:mcmc_3arc_fix} shows the MCMC results for
A\,3556, A\,3558, and A\,3562, and A\,3560 respectively.
Table~\ref{tab.mcmc_fixpos} summarizes the results. For A\,3562, the
estimation for the concentration parameter did not converge,
presumably because the off-centring between the position of the BCG
and the weak lensing centre or large noises in galaxy shapes.

\begin{figure*}
\subfigure{\includegraphics[width=0.8\columnwidth]{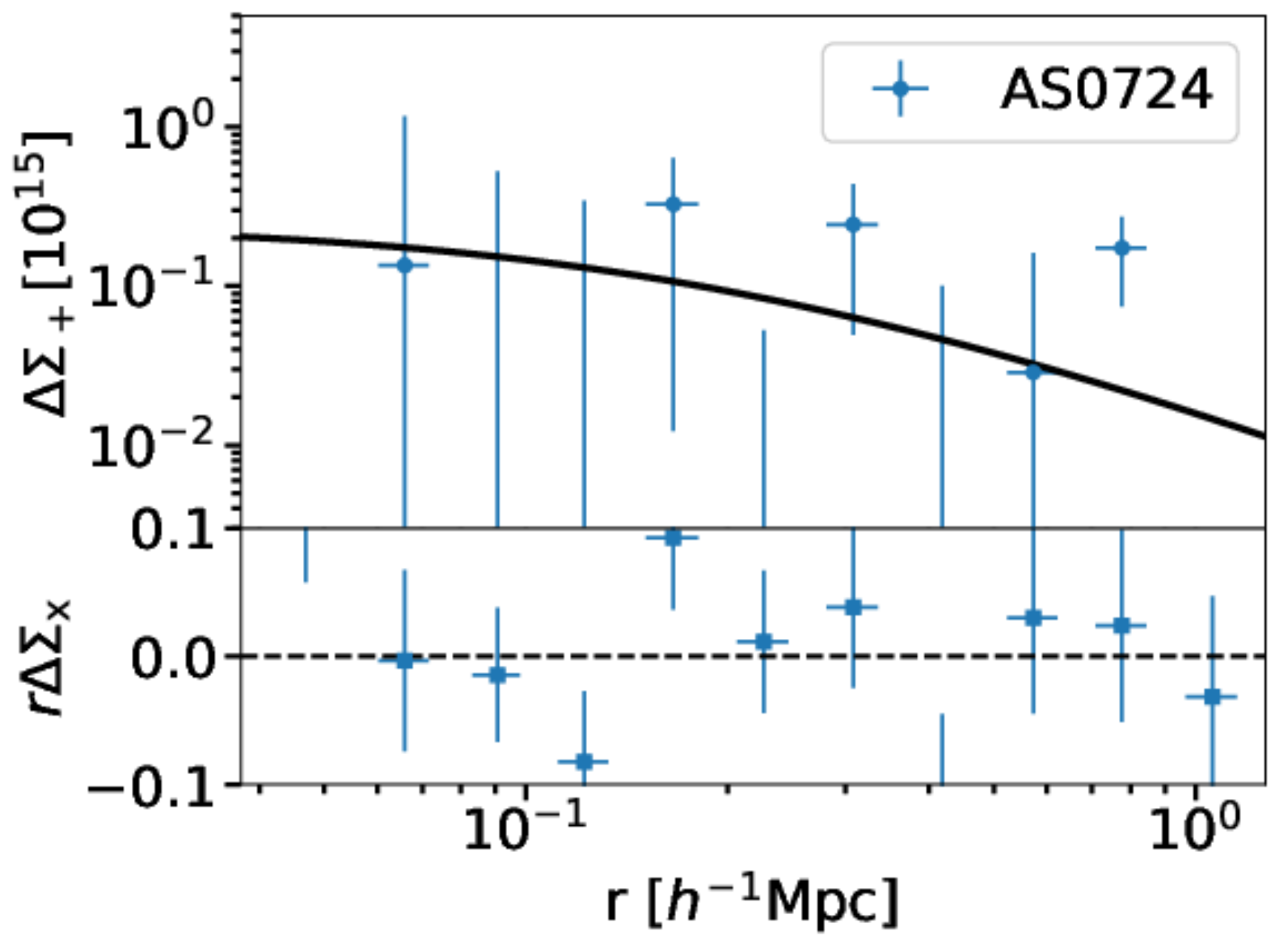}}
\subfigure{\includegraphics[width=0.8\columnwidth]{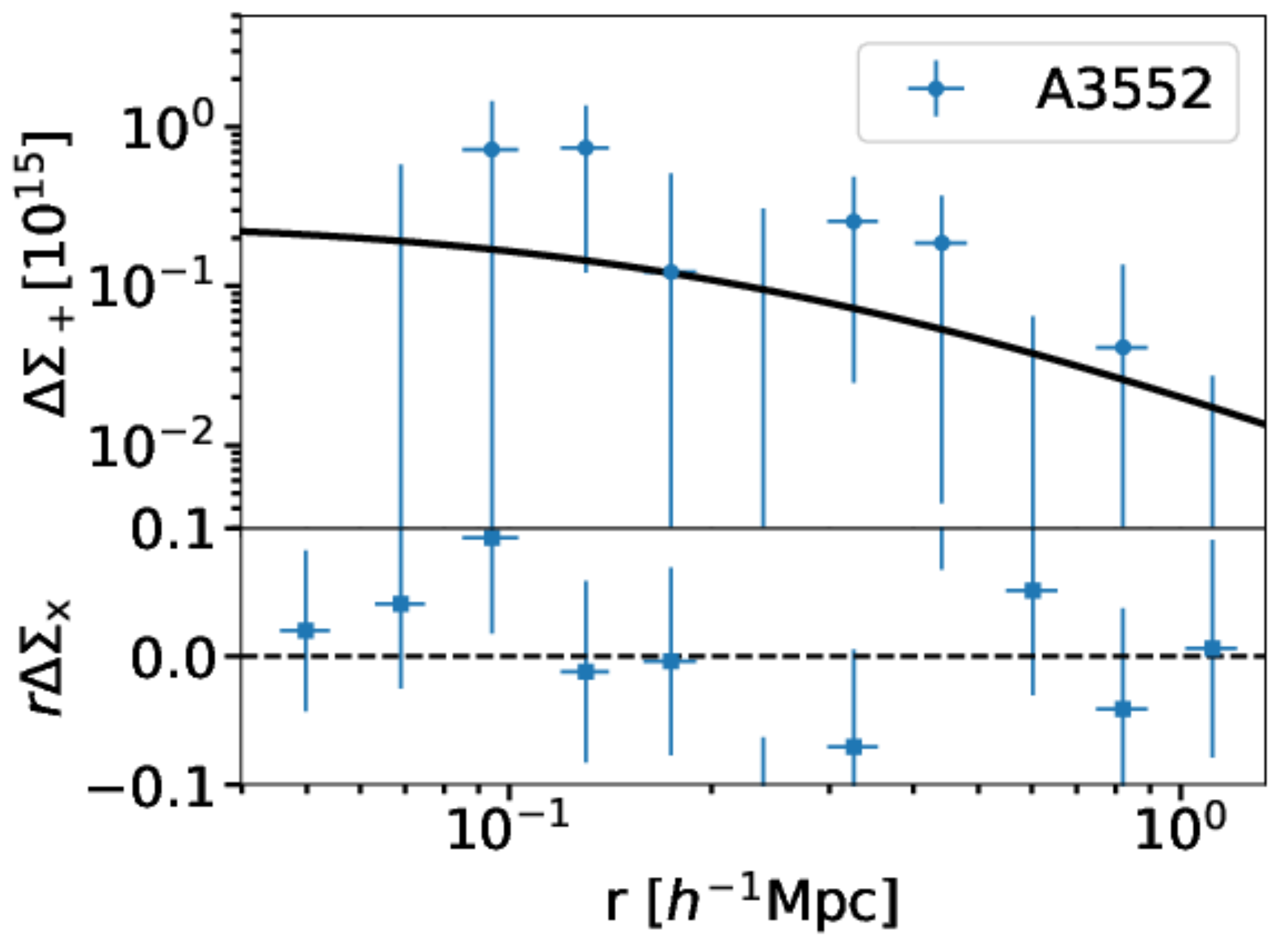}}
\subfigure{\includegraphics[width=0.8\columnwidth]{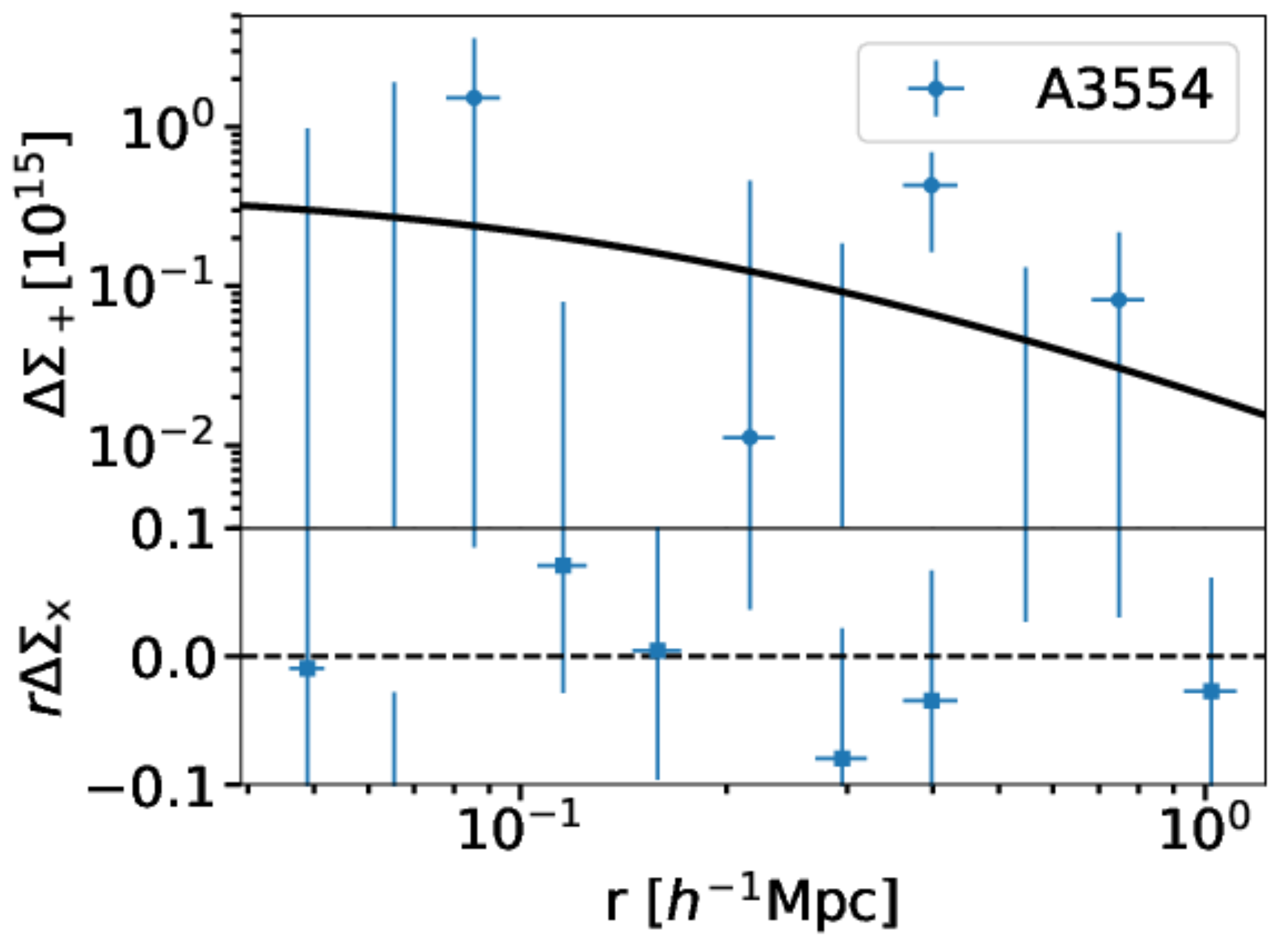}}
\subfigure{\includegraphics[width=0.8\columnwidth]{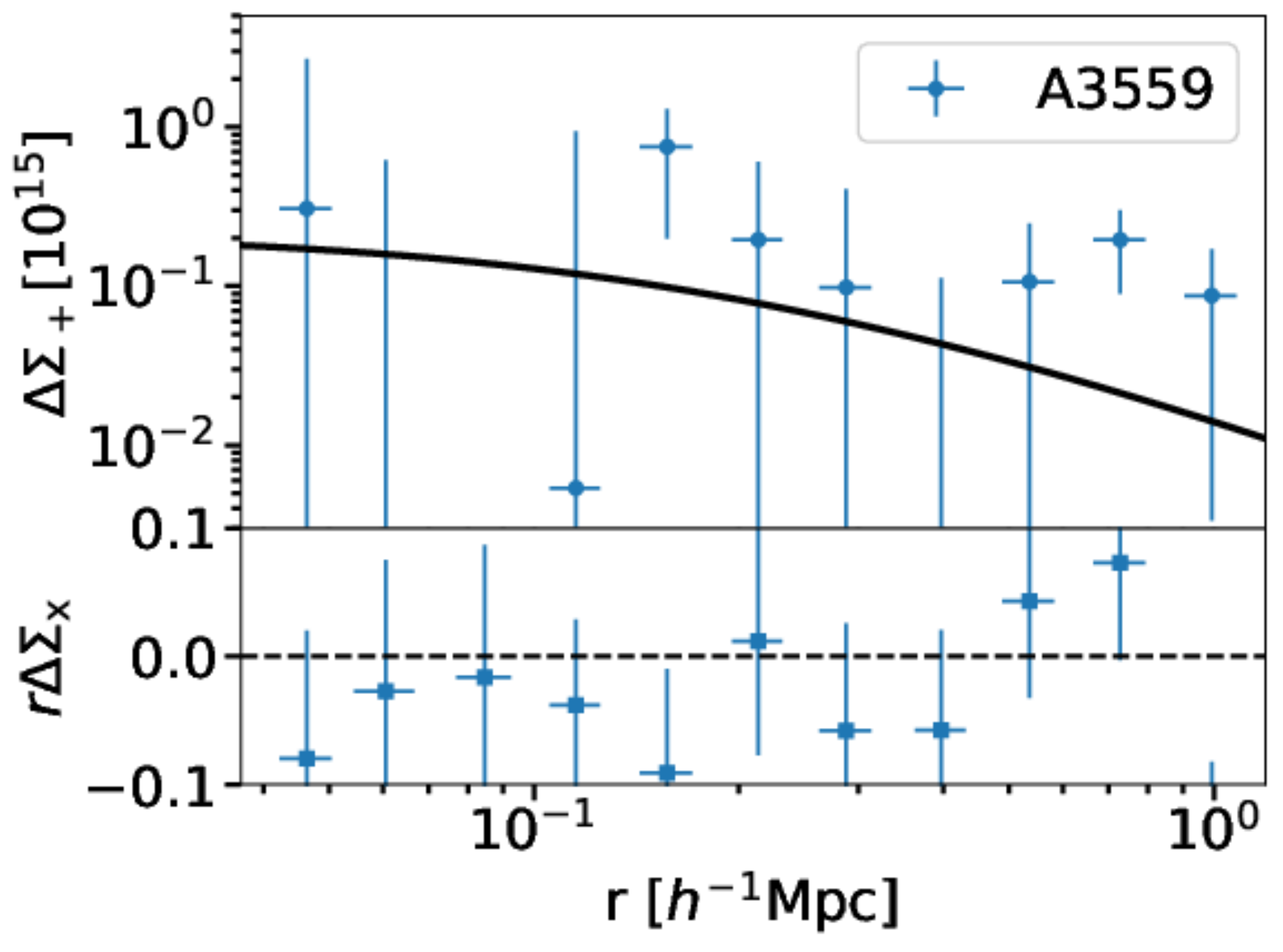}}
\subfigure{\includegraphics[width=0.8\columnwidth]{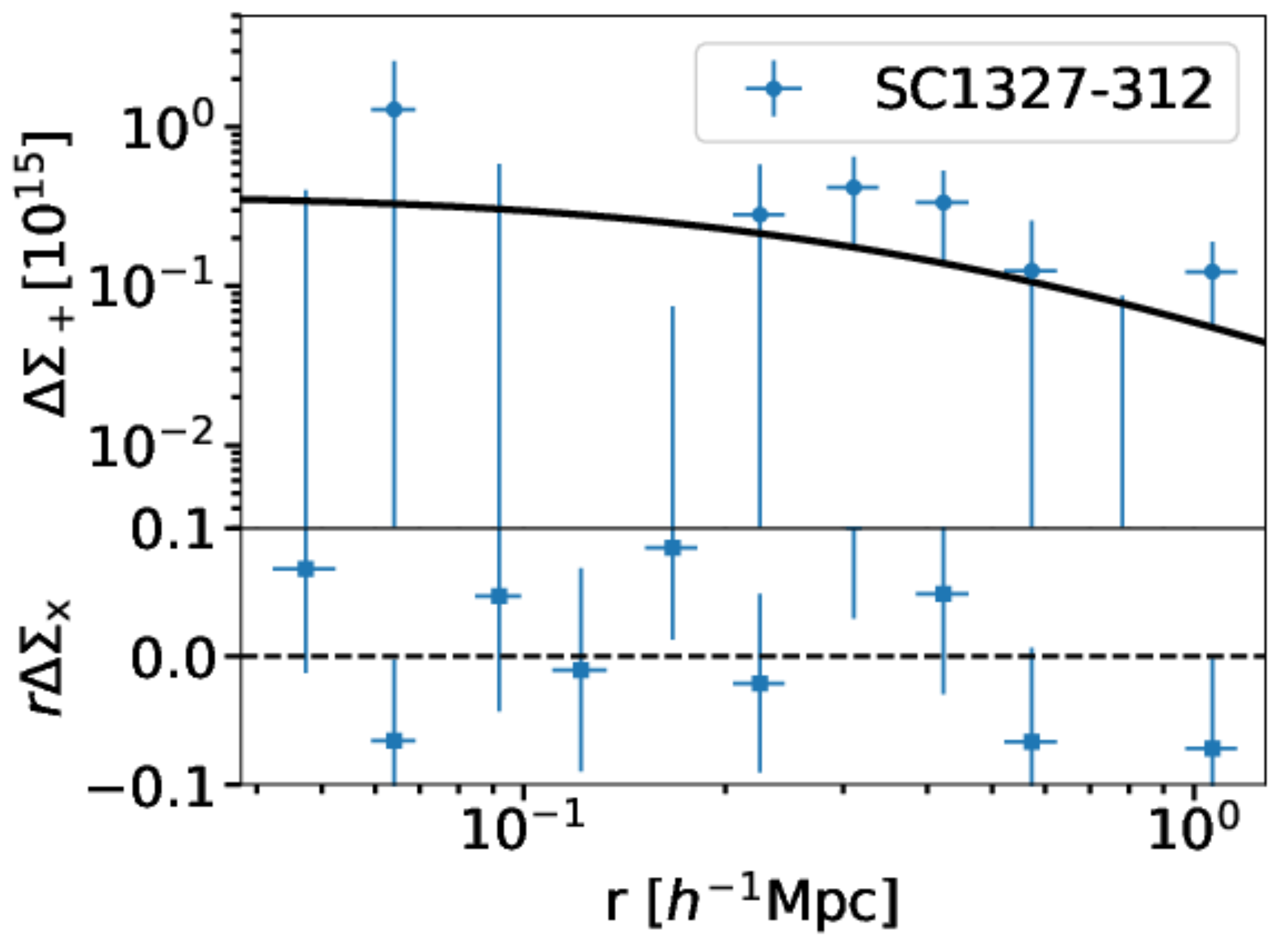}}
\subfigure{\includegraphics[width=0.8\columnwidth]{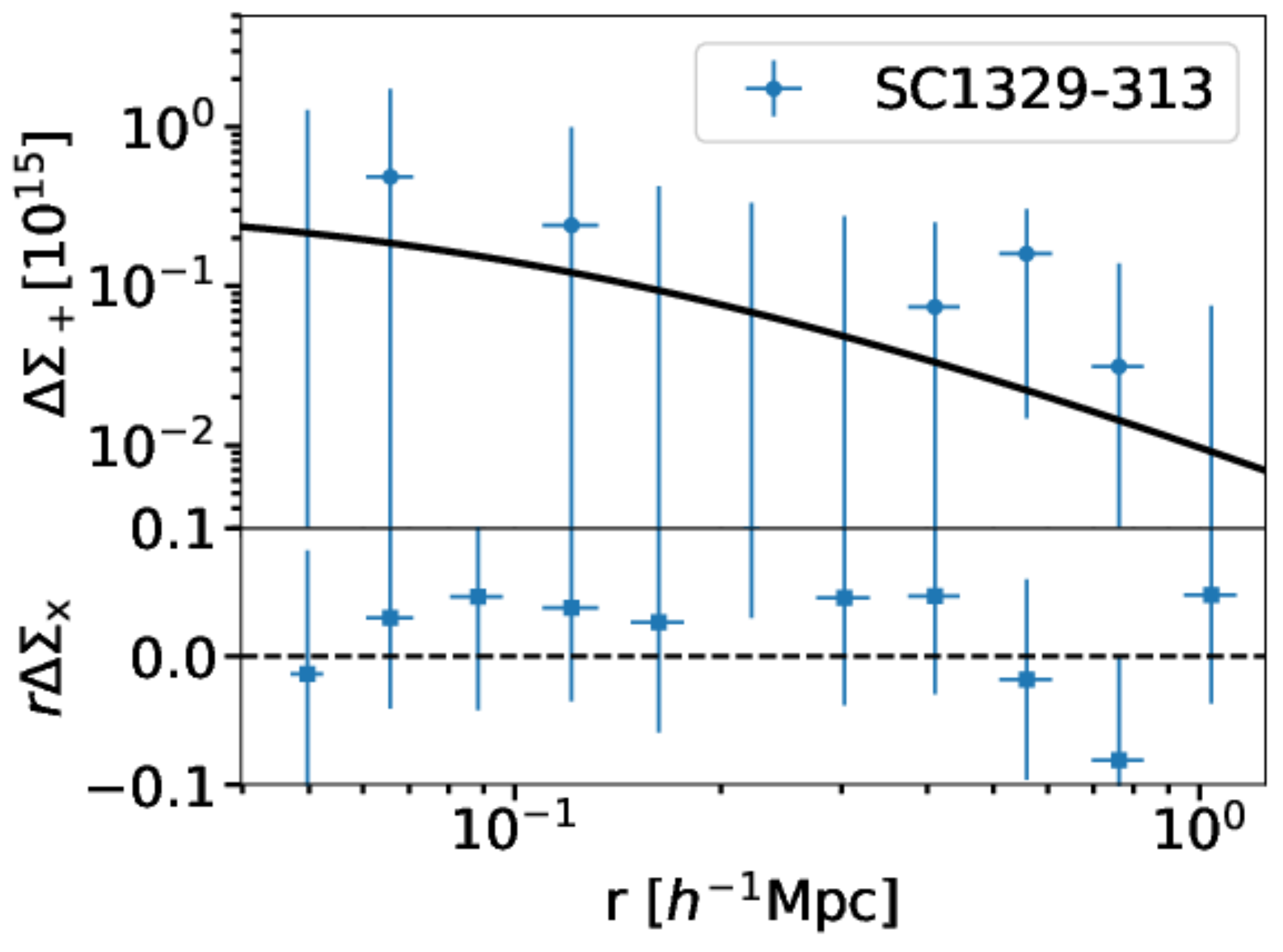}}
\caption{Tangential and cross shear profiles for the low mass clusters.
The horizontal axis shows the distance from the BCG.
The vertical axes for upper and lower panels show a tangential and cross shear profiles, respectively.
The best-fitted NFW profiles are plotted as solid lines. }
\label{fig.ind_shear}
\end{figure*}

For the clusters with lower masses (AS\,0724, A\,3552, A\,3554,
A\,3559, SC\,1327-312 and SC\,1329-313), we fitted their tangential
shear profiles with the NFW profile in 11 annular bins from the BCG
position out to 30\,arcmin. 
In order to check the effects of the bin size on the obtained parameters, we increased and decreased the number of bins and fitted their profiles.
We find that the results turn out to be consistent within the errors. 
The derived best-fit parameters are summarized in Table~\ref{tab.ind_fit}. Due to the low lensing
signal, the fit does not converge for
AS\,0726. Fig.~\ref{fig.ind_shear} shows the measured tangential and
cross shear profiles with the best fitted NFW profile for each
cluster. 

For comparison and to investigate the average properties of the
low-mass clusters, 
we also fitted the tangential shear profile obtained by
stacking the shear profiles of the individual clusters weighted with
the errors.
In the stacking analysis, we included the profile of the cluster A\,0726 and stacked the seven low mass clusters.

\begin{figure}
    \centering
    \includegraphics[width=1\columnwidth]{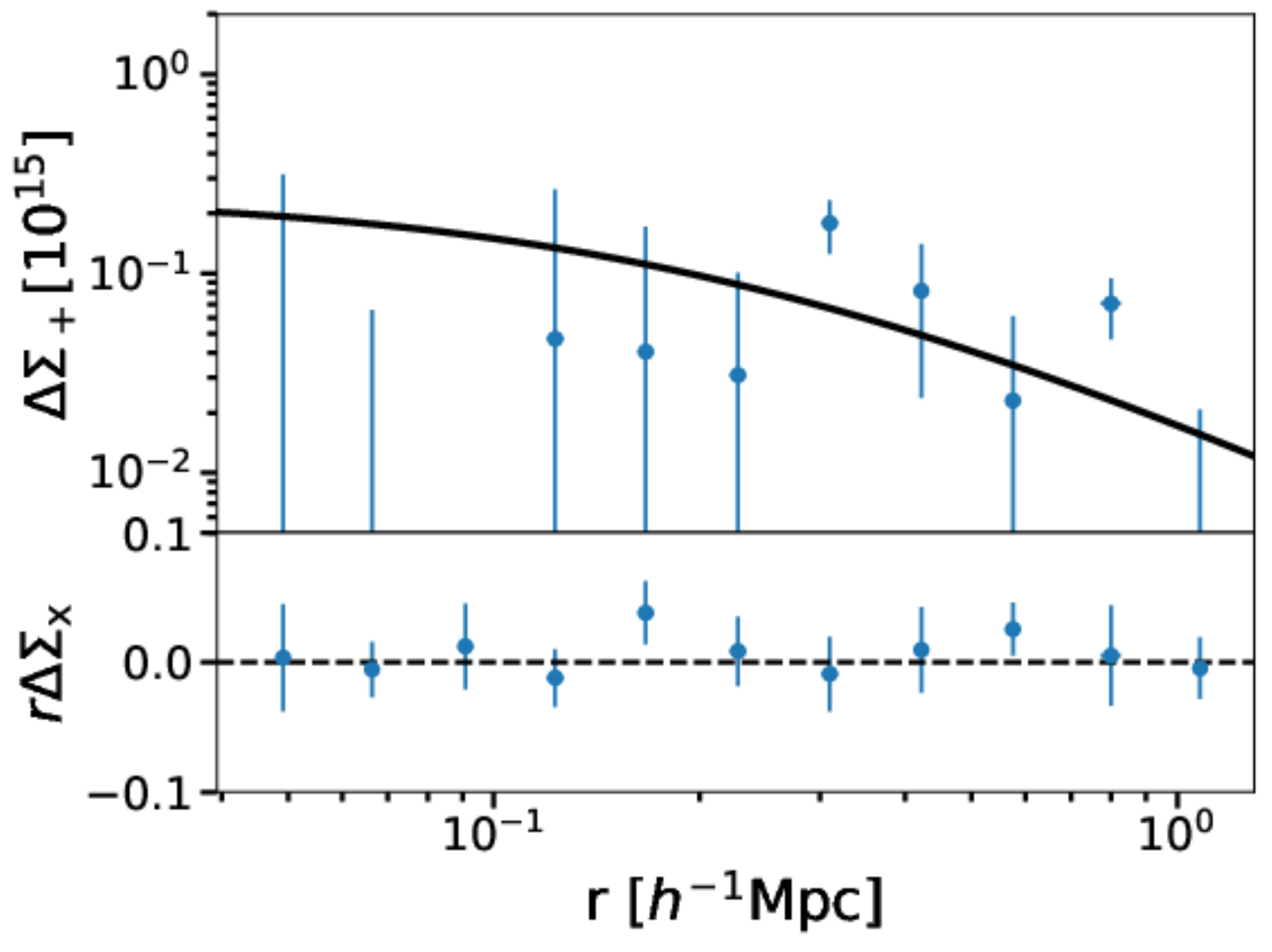}
    \caption{The stacked shear profile for the seven low mass clusters.
    The horizontal axis shows the distance from the BCGs.
    The upper and lower panels show the tangential and cross shear, respectively.
    The points show the results obtained from the observation.
    The solid line shows the best fitted result with the NFW profile.
    }
    \label{fig:stack_low}
\end{figure}

Fig.~\ref{fig:stack_low} shows the stacked shear profile with the
best fitted profile.  The estimated parameters are
$M_{200}=(2.45^{+3.12})\times10^{13}h^{-1}{\rm M}_\odot$ and
$c=3.68^{+6.44}$. The large error is mainly caused by the small number of
background galaxies.


\begin{table*}
\caption{Fitting results of the massive clusters.  The masses and
  concentration parameters are estimated by using the two-dimensional
  shear map with the pixel size of 3 arcmin. $1\sigma$ uncertainties
  are reported.  Column (1): cluster name, Column (2): R.A. of BCG
  [degrees], Column (3): Dec. of BCG [degrees], Column (4): log($M_{200}
  [h^{-1}M_\odot]$), Column (5): concentration parameter.}
  \begin{tabular}{|c|c|c|c|c|c|}
    & Cluster name & R.A. & Dec. & $M_{200}[10^{14}h^{-1}M_\odot]$ & $c$\\\hline\hline
    & A\,3556  & 201.028071 & -31.669883  & $1.62^{+1.40}_{-1.08}$ & $2.35^{+3.75}_{-0.47}$\\
    & A\,3558  & 201.986930 & -31.495891 & $4.47^{+2.78}_{-2.38}$  & $2.63^{+1.23}_{-0.55}$\\
    & A\,3560  & 203.107375 & -33.135833 & $2.75^{+3.85}_{-2.25}$  & $2.49^{+4.72}_{-0.68}$\\
    & A\,3562  & 203.394788 & -31.672261 & $2.04^{+2.74}_{-1.78}$  &  \\\hline
    \end{tabular}
\label{tab.mcmc_fixpos}
\end{table*}

\begin{table*}
\caption{Fitting results for the tangential shear profiles.
Column~(1): cluster name, Column~(2): R.A. of BCG [degrees], Column~(3): Dec. of BCG [degrees], Column~(4): $M_{200} [h^{-1}M_\odot]$, Column~(5): concentration parameter, Column~(6): chi-square}
  \begin{tabular}{|l|c|c|c|c|c|}
    Cluster name & R.A.  & Dec. & $M_{200}[10^{13}h^{-1}M_\odot]$ & $c$ & $\chi^2/$d.o.f \\\hline\hline
    AS\,0724     & 198.247761 & -33.0026810 & $2.34^{+8.09}$     &  $3.90^{+18.84}$  &   9.18/9 \\
    A\,3552      & 199.729591 & -31.8175762 & $3.01^{+17.22}$    &  $3.73^{+24.51}$  &  14.33/9 \\
    A\,3554      & 199.881979 & -33.4881320 & $4.29^{+6.49}$     &  $5.67^{+2.34}$  &  5.97/9 \\
    A\,3559      & 202.462143 & -29.5143931 & $1.79^{+14.92}$    &  $3.55^{+27.45}$  &  8.38/9 \\
    SC\,1327-312 & 202.367236 & -31.5512698 & $4.31^{+49.25}$    &  $2.39^{+4.66}$   &   11.21/9 \\
    SC\,1329-313 & 202.864741 & -31.8206321 & $1.79^{+2.93}$     &  $6.35^{+21.76}$  &   3.12/9 \\ \hline
    \end{tabular}
\label{tab.ind_fit}
\end{table*}

\begin{figure}
\includegraphics[width=1.0\columnwidth]{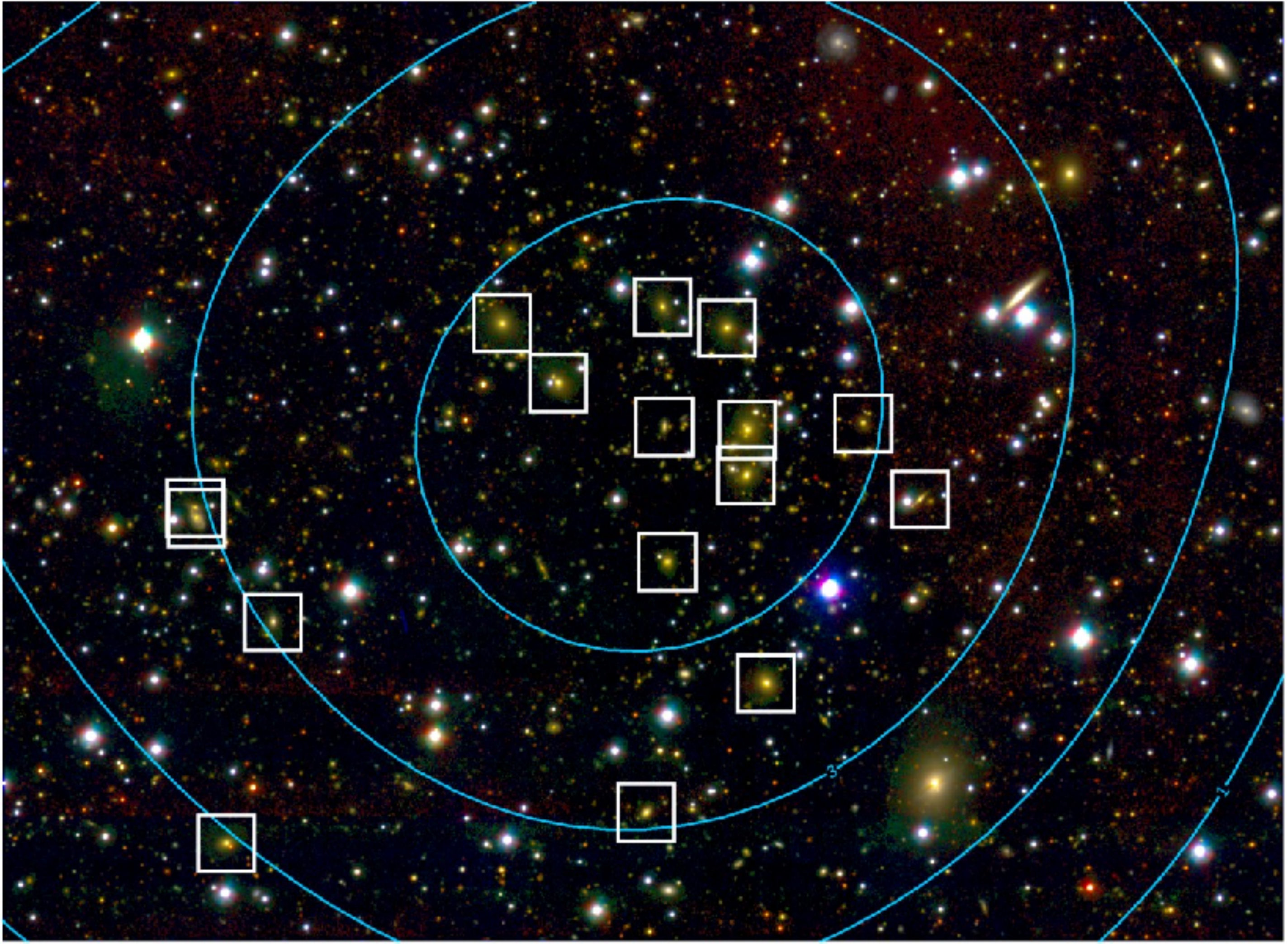}
\includegraphics[width=1.0\columnwidth]{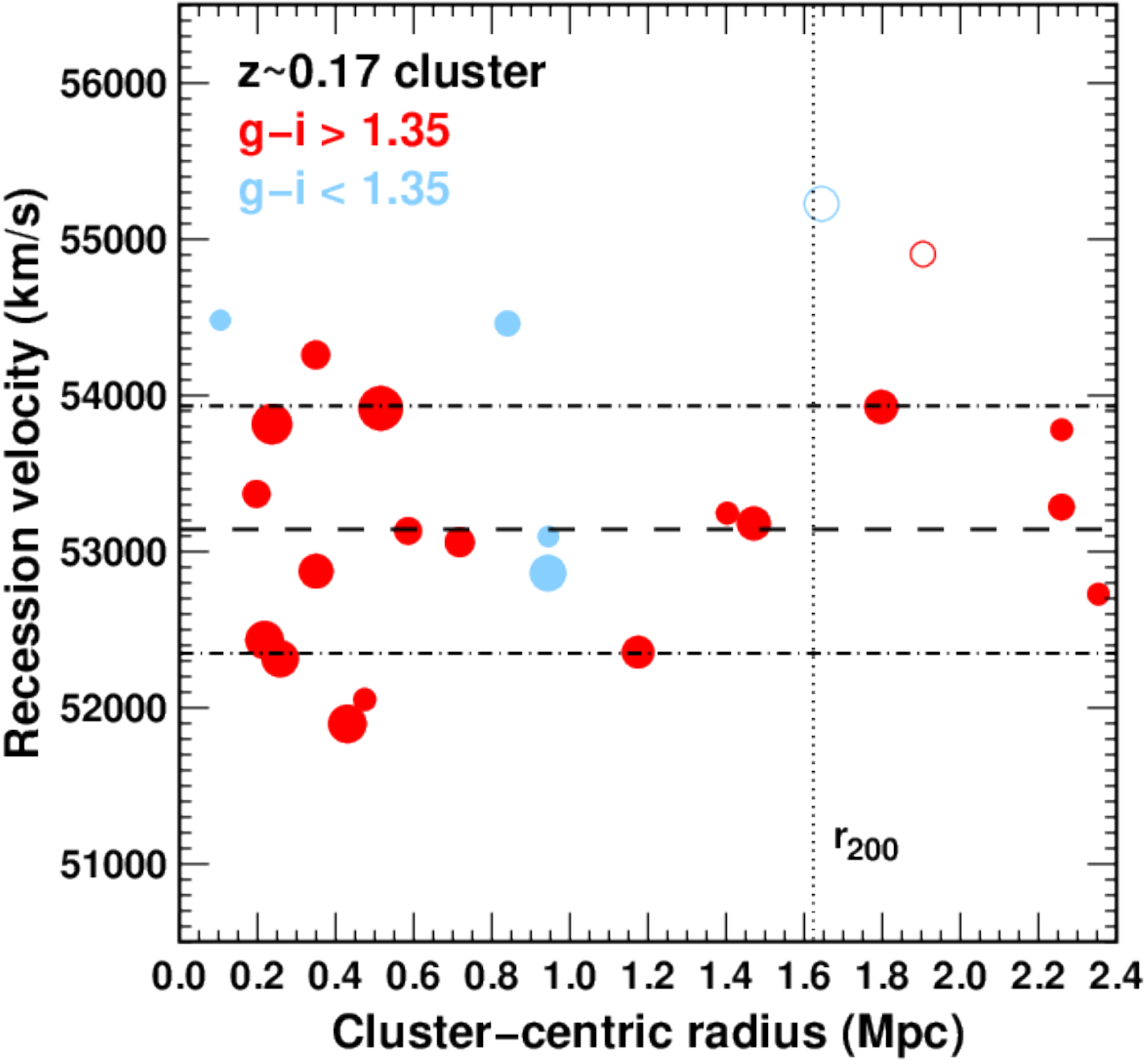}
\caption{\textit{Upper panel}: $gri$ colour composite image centred on
  the WL detection (blue contours) of a background cluster at redshift
  $z\sim0.17$.  The blue contours indicate the WL peak. \textit{Lower
    panel}: Distribution of cluster members of the background cluster
  (solid dots, sizes scale with $i$-band luminosity) and non-members
  (open circles) in the caustic diagram; line-of-sight velocity,$cz$,
  versus projected distance from the centre of the WL peak.  The
  dashed and dot-dashed lines indicate the central velocity and
  1$\sigma$ velocity dispersion of member galaxies within $r_{200}$
  (vertical dotted line).}
\label{fig.bkg_cluster}
\end{figure}

\begin{figure}
    \centering
    \includegraphics[width=1\columnwidth]{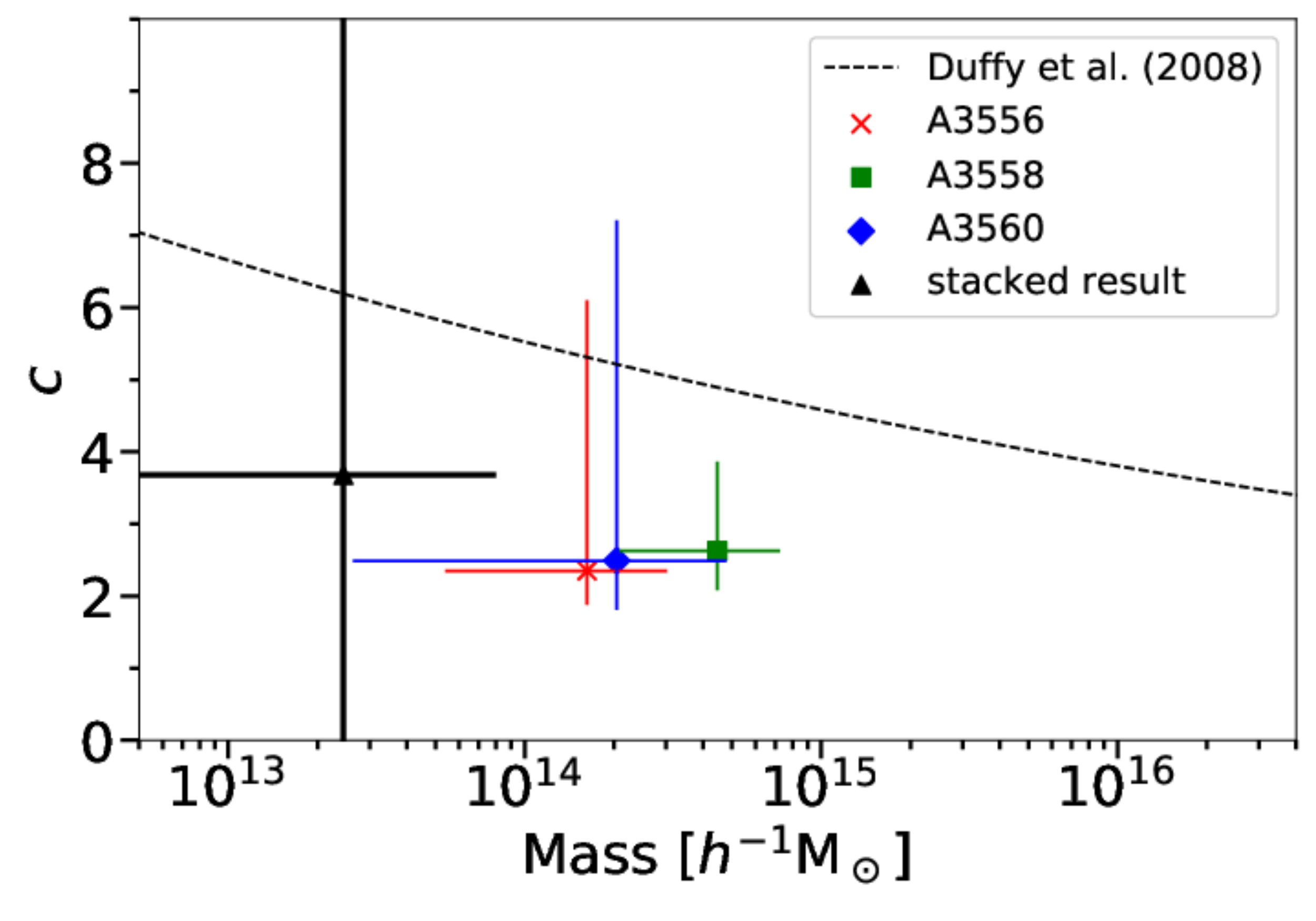}
    \caption{ 
      Comparison of $c-M$ relation between our result and the
      simulations of \citet{2008MNRAS.390L..64D}. 
      The horizontal axis shows $M_{200}$.
      The vertical axis shows concentration parameter.
      The dashed line shows the simulation result.  
      The error bars show $1\sigma$ uncertainties. } 
    \label{fig:c-m_relation}
\end{figure}

\section{discussion}
\label{sec:dis}

\subsection{WL mass map revealing the supercluster and background structures}

In Fig.~\ref{fig.WL_dens} we showed the number density map of
supercluster galaxies across the ShaSS as derived by
\citet{2018MNRAS.481.1055H}, superimposed with the contours of the WL
mass maps obtained from the $r$- (upper panel) and $g$-band (lower
panel) imaging. As already noted, both the WL maps show an overall
agreement with the structure as traced by the galaxy number density,
albeit with a few divergences due to either the different depths or
distortions affecting certain fields. Both the WL maps well trace the
supercluster core including the five clusters. In particular, the WL
map obtained with $r$-band imaging shows a continuous structure
between the five clusters at the $1\sigma$ level (upper panel in
Fig.~\ref{fig.WL_dens}).

\citet{2015MNRAS.446..803M} and \citet{2018MNRAS.481.1055H} found a
filament in the galaxy distribution which connects A\,3559 to the centre of the supercluster.
The density contrast of filaments being so low \citep{2013A&A...559A.112M,
  2014MNRAS.441..745H}, it is difficult to significantly detect such a
structure in our analysis, however the WL contours have a trend to follow the filament connecting A\,3559 to the supercluster core.

The WL mass maps also reveal a number of peaks that do not appear to
be associated with any known cluster within the SSC. These peaks could
instead be due to clusters located behind the SSC.

We take advantage of our extensive spectroscopic coverage of the
entire ShaSS region to investigate the nature of these peaks. 
The $r$-band WL mass reconstruction shows a $\sim4\sigma$ mass peak
at RA=$202{\fdg}1$, Dec=-$32{\fdg}7$ that is not located near to any of the SSC
clusters or any plausible grouping of SSC member galaxies.  An
examination of the redshifts of galaxies in the immediate vicinity of
the WL peak reveals that the nine nearest galaxies
($d{<}160^{\prime\prime}$) all have $z\sim0.177$, with a further nine
$z \sim 0.177$ galaxies located within 8\,arcmin. The $gri$ colour
composite image of the region (Fig.~\ref{fig.bkg_cluster}) shows
that the WL peak (blue contours) is centred on this dense
concentration of red sequence galaxies, which are confirmed to lie
within the redshift range $0.173<z<0.182$ (white squares).  The lower
panel shows the corresponding distribution of galaxies in the caustic
diagram, plotting recession velocity ($V_\mathrm{h}$) versus projected
distance from cluster centre (defined by the peak in the WL map),
confirming that cluster membership is well-defined for this
$z\sim0.177$ system, with noticeable velocity gaps above
54\,500\,km\,s$^{-1}$ and below 51\,700\,km\,s$^{-1}$ where no
galaxies are seen within 1.5\,Mpc of the cluster centre.  The biweight
estimator \citep{1990AJ....100...32B} was used to derive a central
recession velocity of 51\,155\,km\,s$^{-1}$ ($z=0.1773$) for the
cluster and a velocity dispersion of $652{\pm}84$\,km\,s$^{-1}$, based
on 18 member galaxies within $r_{200}$ (1.62\,Mpc), where $r_{200}$ was
iteratively estimated from the $\sigma_{\nu}$ as in
\citet{2018MNRAS.481.1055H}. This implies a mass
$M_{200}{=}5.7{\times}10^{14}{\rm M}_{\odot}$, comparable to that of
Abell\,3556.

The nature of the 5$\sigma$ mass peak located at RA=$198{\fdg}8$,
Dec=-$33{\fdg}03$ is less clear. It is located 
20$^{\prime}$ South of the nearest cluster AS\,0724 within the Shapley
supercluster, but also appears offset by 20$^{\prime}$ West from the
extended structure at $z{\sim}0.10$ that runs up the Western boundary
of the VST survey region \citep{2014MNRAS.445.4073C,
  2018MNRAS.481.1055H}. The most likely correspondence appears to be
with groups of galaxies at $z\sim0.1$ to $z\sim0.3$.

\subsection{The $c-M$ relation of ShaSS massive clusters}

Fig.~\ref{fig:c-m_relation} shows a comparison of the concentration
parameters obtained from our fitting results and the $c-M$ relation
derived by the simulations of \citet{2008MNRAS.390L..64D}. For the
analytical model, the mean redshift of the clusters were adopted. 

\begin{figure*}
\subfigure{\includegraphics[width=1.0\columnwidth]{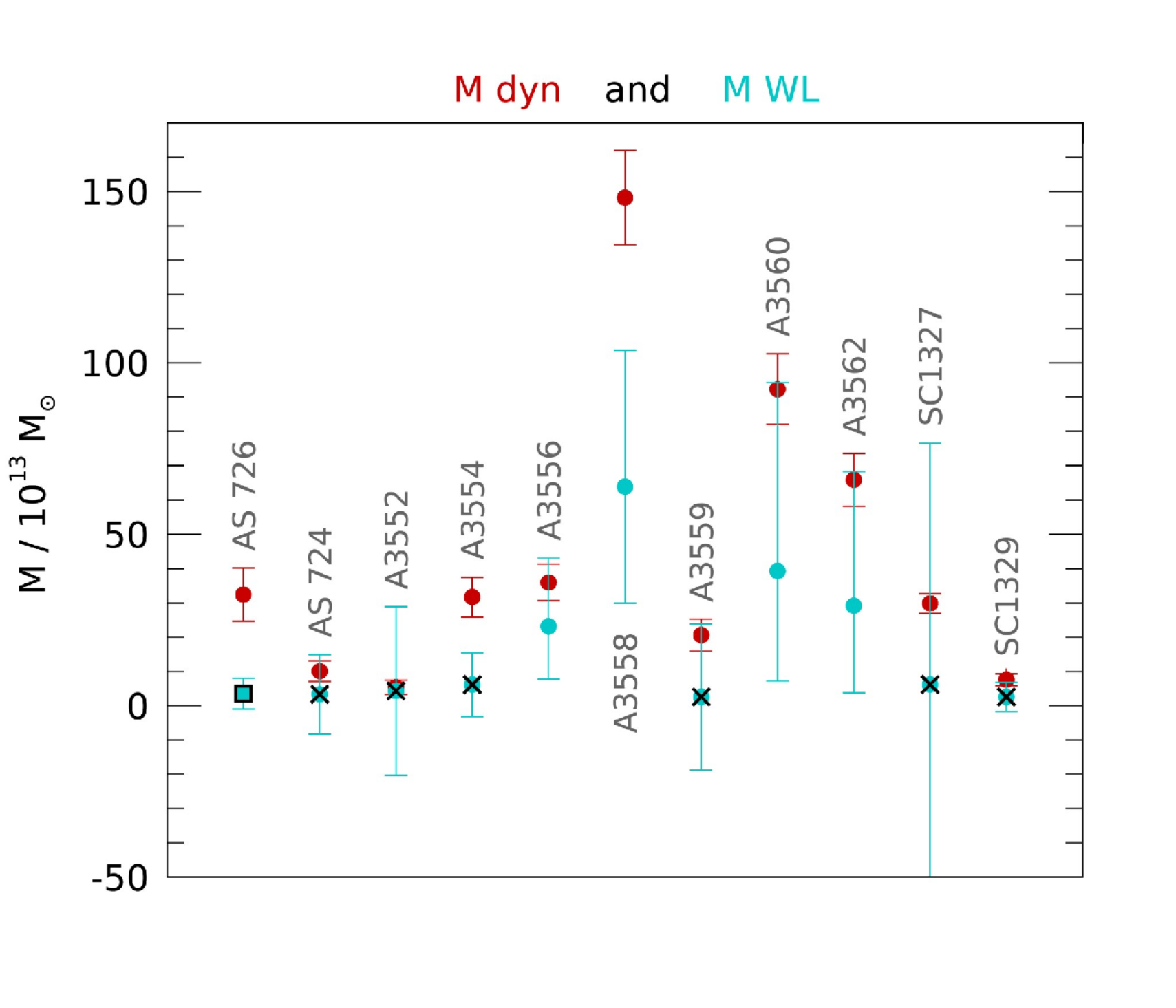}}
\subfigure{\includegraphics[width=1.0\columnwidth]{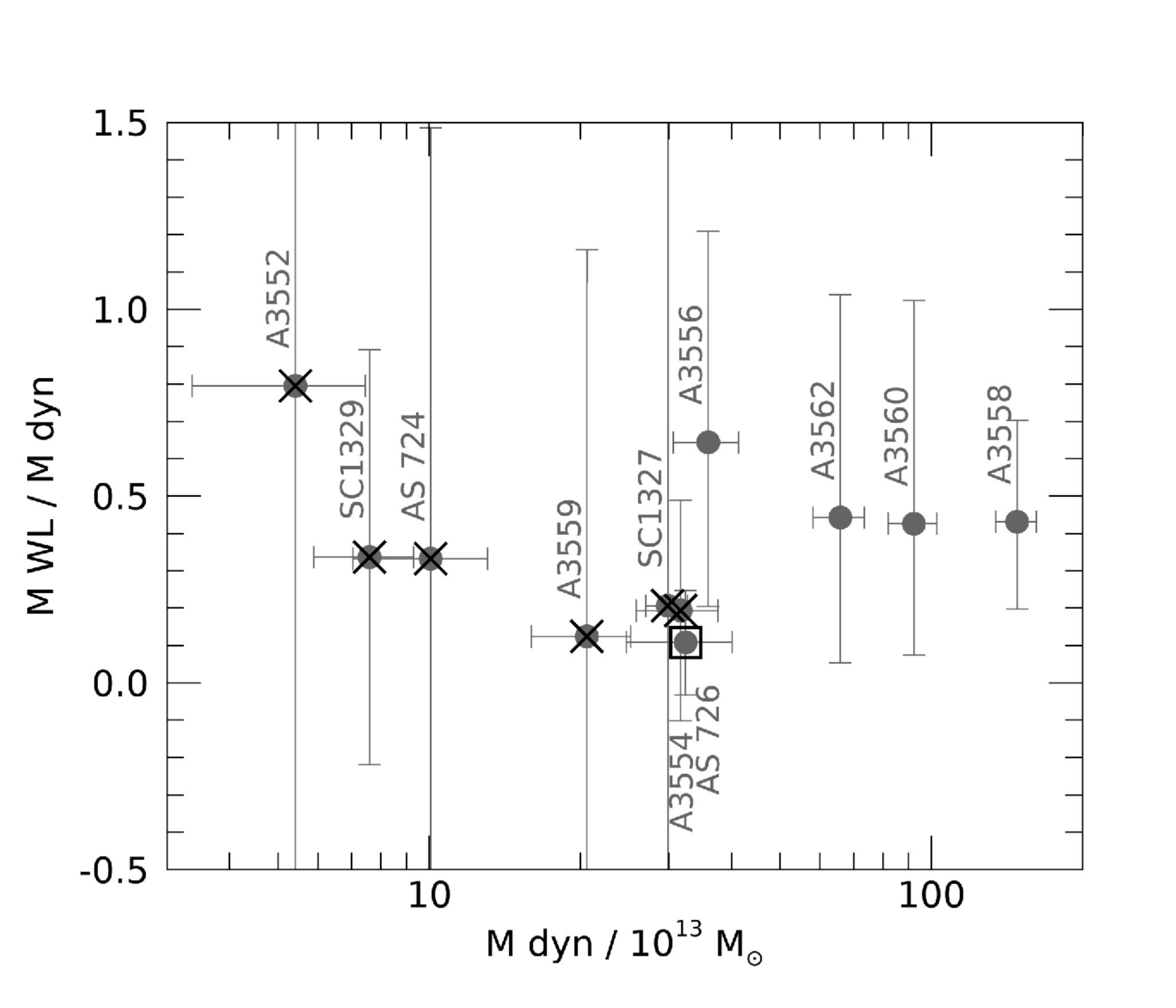}}
\caption{{\it Left panel}. The WL-derived masses (M$_\mathrm{WL}$, cyan
  circles) for the 11 clusters are compared with the with the
  dynamical masses (M$_\mathrm{dyn}$, red circles). Black crosses denote the
  low-mass clusters (see Table~\ref{tab.clus_prop}). Open black square denotes
  AS0726 for which the average mass for the low-mass clusters is
  adopted. {\it Right panel}. The ratio of WL-derived and dynamical
  masses as function of the dynamical mass for the 11 clusters. The
  1$\sigma$ error bar is indicated in both panels.}
\label{fig.masses}
\end{figure*}

Although our results turn out to be consistent with the simulations
within $2\sigma$, the derived $c-M$ relations tend to be lower.  While
the results have large uncertainties, they could be explained by the
dynamical state of the clusters.  Previous studies showed that values
of concentration parameters for unrelaxed clusters are lower than
those for relaxed clusters of the same mass
\citep{2007MNRAS.381.1450N, 2013ApJ...766...32B, 2018ApJ...859...55C,
  2019PASJ...71...79O} and unrelaxed clusters having $\sim30\%$ lower
values of the concentration parameter in our mass range. Based on
hierarchical structure formation, the high-overdensity environment of
superclusters are more recently formed than normal clusters, so that
the concentration of their components are expected to be
lower. Therefore, the dynamical state or formation epoch of the
constituent clusters could explain the smaller values of the
concentration parameters in our results.

\subsection{Cluster masses}

In Fig.~\ref{fig.masses}, we compared the WL-derived masses (M$_\mathrm{WL}$)
with those derived from the dynamical analysis(M$_\mathrm{dyn}$) and listed in Table~\ref{tab.clus_prop}. 
The dynamical masses were computed under the assumption of the singular isothermal model and velocity dispersions \citep{2018MNRAS.481.1055H}.
We obtained the WL masses for the four massive clusters (A\,3556, A\,3558, A\,3560 and A\,3562) with the MCMC method.
On the other hand, we only gave the upper limit of the masses for 6 out of 7 low-mass clusters due to the large shape noise in the WL analysis.
In the figure, we labeled the low-mass clusters as black crosses, AS\,0726 as black box and the four massive clusters as cyan circles.
For AS\,0726, we only indicatively adopted the average mass for the low-mass clusters derived using the stacked shear profile.
We notice that (i) the masses obtained from the dynamical and WL analysis are consistent within 1$\sigma$ for all clusters except A\,3554, A\,3558 and AS\,0726; (ii) the dynamical masses turn out to be systematically higher than the WL- derived ones. 
Previous studies showed that WL masses obtained from the tangential shear fitting were biased low up to 10 percent with a scatter of $\sim25$ percent \citep{2005ApJ...632..841O, 2011MNRAS.416.3187S, 2015MNRAS.450.3633S}.
The main source of the bias are due to substructures and triaxiality.
When a cluster whose major axis is perpendicular to the line of sight, i.e elongated in the sky-plane, a mass obtained with the spherical NFW profile is typically underestimated.
The presence of substructures around clusters and uncorrelated large-scale structures along the line of sight also generate the biases for the estimation of the WL masses \citep{2010A&A...514A..93M, 2011ApJ...740...25B, 2012MNRAS.426.1558G, 2014MNRAS.440.1899G}.
In addtion, we point out that the dynamical mass of AS\,0726 is actually an upper limit since the velocity distribution of member galaxies is strongly bimodal, and certainly not Gaussian \citep[see Fig.~14 in][]{2018MNRAS.481.1055H}. 
This system probably consists of two groups with velocity dispersions $\sim 300$\,km\,s$^{-1}$ rather than one single system with $\sigma \sim 600$\,km\,s$^{-1}$. 
This would reduce the mass estimate by a factor 4. 
The complex structure of A\,3558 and, in general, the dynamical activity in the SSC core \citep[see][]{1998MNRAS.300..589B,EBd00,Finoguenov_2004,RGM07} may
explain the systematic differences between the two mass determinations quoted above, since the virial mass tends to overestimate the mass of unrelaxed clusters. 
Moreover, \citet{2007A&A...463..839R} studied A\,3558 with a X-ray observation and showed the possibility of the presence of substructures along the line of sight, which interact with the cluster. 
Since such structures along the line of sight broaden velocity distribution, masses obtained from dynamical analysis can be overestimated up to 100\% \citep{2010PASJ...62..951T, 2019SSRv..215...25P}. 
Fig.~14 in \citet{2018MNRAS.481.1055H} showed the broad velocity distribution of galaxies for the cluster.
This indicates the presence of the substructures along the line of sight and the possibility for overestimating the dynamical mass.  
A\,3554 also shows strong substructures in the velocity distribution diagram. 

The WL and dynamical mass estimates for the massive clusters are actually consistent within 1$\sigma$ for three clusters and within 2$\sigma$ for A3558, even though we used different measurement techniques and different mass models.
Thus, the total mass of the 4 massive clusters does not dramatically differ between the two approaches. 
We found their total masses $\mathcal{M}_{\rm 4cl,dyn} = (3.42\pm 0.20)\times 10^{15}{\rm M}_{\odot}$ and $\mathcal{M}_{\rm 4cl,WL} = (1.56^{+0.81}_{-0.55})\times10^{15}{\rm M}_{\odot}$ - i.e. only a factor $\sim 2.2$ lower. 
By using the empirical L$_{\rm X}-{\rm M}_{\rm 200}$ scaling relation of \citet{BCC14} (their equation~10), we obtain $\mathcal{M}_{\rm 4cl,X} = (2.23\pm 0.7)\times 10^{15}{\rm M}_{\odot}$.
The error of this estimates reflects the scatter of the mass--X-ray luminosity relation.  
While we could give upper limits of the masses for the low mass clusters, we got the total mass for the 11 clusters $\mathcal{M}_{\rm ShaSS,WL} = (1.84^{+1.13})\times 10^{15}{\rm M}_{\odot}$.
In the calculation, we used the average mass derived from the stacking analysis for the mass of AS\,0726.
The total dynamical mass is $\mathcal{M}_{\rm ShaSS,dyn} = (4.80\pm 2.3)\times 10^{15}{\rm M}_{\odot}$.

\citet{1995AJ....110..463Q} estimated a mass for the whole Shapley
supercluster in the range $7\times 10^{15} - 7\times 10^{16}{\rm
  M}_{\odot}$\footnote{In the following all the masses are converted
  to H$_0 = 70$\,km\, s$^{-1}$\,Mpc$^{-1}$.}. \citet{RMP06} identified
122 galaxy systems across a $12\times 15$ degrees region centred in
the SSC core and estimated their individual masses which summed up
result in a total mass of $\mathcal{M}_{SSC} = 6.88\times 10^{15}{\rm
  M}_{\odot}$. From the observed galaxy overdensity in a 285\, deg$^2$
region, \citet{2006A&A...447..133P} evaluated for the supercluster a
total mass of $\mathcal{M}_{SSC} = 7\times 10^{16}{\rm
  M}_{\odot}$. They also claimed that the result of \citet{RMP06} is a
lower limit for the SSC mass.

All these mass estimates refered to the whole supercluster, on the other
hand ShaSS covers 260\,Mpc$^2$ centred on the core. Applying a
spherical collapse model, \citet{2000AJ....120..523R} found that the
SSC is gravitationally collapsing at least in its central region
within a radius of 8\,$h^{-1}$\,Mpc, centred on A\,3558 including 11 clusters; the
very inner region, associated with the massive clusters, is likely in
the final stages of collapse. The mass within this radius was found to
be $\mathcal{M}_\mathrm{SSC} \sim 1.4\times 10^{16}{\rm M}_{\odot}$. 
On the other end, they same authors
using the heuristic escape-velocity methods of \citet{Diaferio_1997}
obtained a mass estimate of $\mathcal{M}_\mathrm{SSC} \sim 2.9\times
10^{15}{\rm M}_{\odot}$. 
This value is consistent within the errors with both our measurements of
$\mathcal{M}_{\rm ShaSS,dyn}$ and $\mathcal{M}_{\rm ShaSS,WL}$.

\section{Summary and conclusions}
\label{sec:con}
With the aim to investigate the mass distribution and thus to trace
the environment in the centre of the Shapley supercluster (at
$z\sim0.048$), we have conducted the first weak lensing analysis of a
260\,Mpc$^2$ region around the supercluster core including 11
clusters. This study has taken advantage of the multi-band ($gri$)
optical imaging collected at ESO-VST together with the related
photometric catalogues. These data have allowed us to generate the
galaxy shape catalogues in $g$ and $r$ bands across the whole surveyed
region. In the following the adopted approach is briefly described.

\begin{description}
\item [-] Supercluster members have been selected via photometric redshifts
  and background/foreground galaxies with the
  colour-colour diagram. The average density of the background
  galaxies turned out to be 7 arcmin$^{-2}$.

\item [-] The project mass distributions, i.e. the WL mass maps have
  been derived for both $r$- and $g$-band data. 
  The significance of the mass map has been estimated by randomizing the background galaxy shapes.
  The two maps allowed to double-check the final results especially in one fields affected by peculiar distortion.

\item [-] Concentration parameters and masses were obtained fitting
  with the NFW profile the two-dimensional shear of the massive
  clusters (A\,3556, A\,3558, A\,3560 and A\,3562) and the tangential
  shear of the low-mass clusters (AS\,0724, A\,3552, A\,3554, A\,3559,
  SC\,1327-312 and SC\,1329-313). For the low mass clusters, we have
  also fitted their stacked tangential shear profile estimating their
  average mass and concentration parameter.
\end{description}

Our analysis of the ShaSS provides further evidence of the complex
structure of the system and cluster-cluster interactions, reveals a new
background cluster of galaxies, and provides WL-derived masses for the 11 clusters embedded in a common network.

\begin{description}
\item [-] We have found a tight correlation between WL mass
  distribution and the structure as traced by the galaxy density
  previously derived for the ShaSS. In particular, the WL map in $r$
  band highlights that the SSC core consists of a coherent system and
  shows indications of a filaments connecting the SSC core and A\,3559
  in agreement with our previous study revealing a stream of galaxies
  in the same region.

\item [-] The total WL-derived mass of the 4 massive clusters is
  $\mathcal{M}_\mathrm{ShaSS,WL} = (1.56^{+0.81}_{-0.55})\times 10^{15}{\rm
  M}_{\odot}$, which is consistent with their total dynamical mass. 
  Adding the upper mass limits of the remaining clusters, the total mass is consistent with the total dynamical mass of \citet{2018MNRAS.481.1055H} and with
  \citet{2000AJ....120..523R} who analysed almost the same region of
  the ShaSS.

\item [-] The WL-derived masses are found
  to be systematically lower than the dynamical ones for each cluster,
  although the different estimates are consistent within 1$\sigma$ for
  8 out of 10 clusters. This discrepancy can be explained by the fact
  that in such a perturbed and dynamically-active environment, the
  cluster dynamical mass should be actually considered as an upper
  limit. In fact, the differences between the mass derivations are higher
  in the less relaxed and more substructured clusters. Likewise, the
  $c-M$ relation of ShaSS clusters shows concentration parameters
  typical of unrelaxed clusters.
  
  \item [-] Finally, in the WL mass map, we detect a peak associated to a
  previously unknown background cluster at $z\sim 0.177$ with a
  velocity dispersion of $652{\pm}84$\,km\,s$^{-1}$ (based on 18
  member galaxies within $r_{200}$=1.62\,Mpc) and implying a mass
  $M_{200}{=}5.7{\times}10^{14}{\rm M}_{\odot}$, comparable to that of
  A\,3556.
  
\end{description}

We conclude that the WL and dynamical analyses are complementary and
both essential for a robust characterization of the supercluster
environment allowing us to ascertain the continuity of the SSC structure
around the core and supporting the scenario of an ongoing collapse.

\section*{Acknowledgements}
We would like to thank anonymous referee for giving useful comments and improving our manuscript.
We would like to thank K.Umetsu, I.Chiu and Y.Toba for useful comments
and discussions.  This work is supported by ALMA collaborative science
research project 2018-07A and in part by the Ministry of Science and
Technology of Taiwan (grant MOST 106-2628-M-001-003-MY3) and by
Academia Sinica (grant AS-IA-107-M01). Numerical computations were in part carried out on Cray XC50 at Center for Computational Astrophysics, National Astronomical Observatory of Japan.  Data analyses
were (in part) carried out on common use data analysis computer system
at the Astronomy Data Center, ADC, of the National Astronomical
Observatory of Japan. The optical imaging is collected at the VLT
Survey Telescope using the Italian INAF Guaranteed Time
Observations. PM GB and AM acknowledge financial support from INAF
PRIN-SKA 2017 {\it ESKAPE} (PI L. Hunt).




\bibliographystyle{mnras}
\bibliography{newbib}

\begin{thebibliography}{}
\makeatletter
\relax
\def\mn@urlcharsother{\let\do\@makeother \do\$\do\&\do\#\do\^\do\_\do\%\do\~}
\def\mn@doi{\begingroup\mn@urlcharsother \@ifnextchar [ {\mn@doi@}
  {\mn@doi@[]}}
\def\mn@doi@[#1]#2{\def\@tempa{#1}\ifx\@tempa\@empty \href
  {http://dx.doi.org/#2} {doi:#2}\else \href {http://dx.doi.org/#2} {#1}\fi
  \endgroup}
\def\mn@eprint#1#2{\mn@eprint@#1:#2::\@nil}
\def\mn@eprint@arXiv#1{\href {http://arxiv.org/abs/#1} {{\tt arXiv:#1}}}
\def\mn@eprint@dblp#1{\href {http://dblp.uni-trier.de/rec/bibtex/#1.xml}
  {dblp:#1}}
\def\mn@eprint@#1:#2:#3:#4\@nil{\def\@tempa {#1}\def\@tempb {#2}\def\@tempc
  {#3}\ifx \@tempc \@empty \let \@tempc \@tempb \let \@tempb \@tempa \fi \ifx
  \@tempb \@empty \def\@tempb {arXiv}\fi \@ifundefined
  {mn@eprint@\@tempb}{\@tempb:\@tempc}{\expandafter \expandafter \csname
  mn@eprint@\@tempb\endcsname \expandafter{\@tempc}}}

\bibitem[\protect\citeauthoryear{{Bardelli}, {Pisani}, {Ramella}, {Zucca}  \&
  {Zamorani}}{{Bardelli} et~al.}{1998}]{1998MNRAS.300..589B}
{Bardelli} S.,  {Pisani} A.,  {Ramella} M.,  {Zucca} E.,   {Zamorani} G.,
  1998, \mn@doi [\mnras] {10.1046/j.1365-8711.1998.01930.x}, \href
  {https://ui.adsabs.harvard.edu/abs/1998MNRAS.300..589B} {300, 589}

\bibitem[\protect\citeauthoryear{{Bardelli}, {Zucca}, {Zamorani}, {Moscardini}
  \& {Scaramella}}{{Bardelli} et~al.}{2000}]{2000MNRAS.312..540B}
{Bardelli} S.,  {Zucca} E.,  {Zamorani} G.,  {Moscardini} L.,   {Scaramella}
  R.,  2000, \mn@doi [\mnras] {10.1046/j.1365-8711.2000.03174.x}, \href
  {https://ui.adsabs.harvard.edu/abs/2000MNRAS.312..540B} {312, 540}

\bibitem[\protect\citeauthoryear{{Bartelmann} \& {Schneider}}{{Bartelmann} \&
  {Schneider}}{2001}]{2001PhR...340..291B}
{Bartelmann} M.,  {Schneider} P.,  2001, \mn@doi [Phys.~Rep.]
  {10.1016/S0370-1573(00)00082-X}, \href
  {http://adsabs.harvard.edu/abs/2001PhR...340..291B} {340, 291}

\bibitem[\protect\citeauthoryear{{Becker} \& {Kravtsov}}{{Becker} \&
  {Kravtsov}}{2011}]{2011ApJ...740...25B}
{Becker} M.~R.,  {Kravtsov} A.~V.,  2011, \mn@doi [\apj]
  {10.1088/0004-637X/740/1/25}, \href
  {https://ui.adsabs.harvard.edu/abs/2011ApJ...740...25B} {740, 25}

\bibitem[\protect\citeauthoryear{{Beers}, {Flynn}  \& {Gebhardt}}{{Beers}
  et~al.}{1990}]{1990AJ....100...32B}
{Beers} T.~C.,  {Flynn} K.,   {Gebhardt} K.,  1990, \mn@doi [\aj]
  {10.1086/115487}, \href
  {https://ui.adsabs.harvard.edu/abs/1990AJ....100...32B} {100, 32}

\bibitem[\protect\citeauthoryear{{Bhattacharya}, {Habib}, {Heitmann}  \&
  {Vikhlinin}}{{Bhattacharya} et~al.}{2013}]{2013ApJ...766...32B}
{Bhattacharya} S.,  {Habib} S.,  {Heitmann} K.,   {Vikhlinin} A.,  2013,
  \mn@doi [The Astrophysical Journal] {10.1088/0004-637X/766/1/32}, \href
  {https://ui.adsabs.harvard.edu/abs/2013ApJ...766...32B} {766, 32}

\bibitem[\protect\citeauthoryear{{B\"ohringer, Hans}, {Chon, Gayoung}  \&
  {Collins, Chris A.}}{{B\"ohringer, Hans} et~al.}{2014}]{BCC14}
{B\"ohringer, Hans} {Chon, Gayoung}  {Collins, Chris A.} 2014, \mn@doi [A\&A]
  {10.1051/0004-6361/201323155}, 570, A31

\bibitem[\protect\citeauthoryear{{Bruzual} \& {Charlot}}{{Bruzual} \&
  {Charlot}}{2003}]{2003MNRAS.344.1000B}
{Bruzual} G.,  {Charlot} S.,  2003, \mn@doi [\mnras]
  {10.1046/j.1365-8711.2003.06897.x}, \href
  {https://ui.adsabs.harvard.edu/abs/2003MNRAS.344.1000B} {344, 1000}

\bibitem[\protect\citeauthoryear{{Bullock}, {Kolatt}, {Sigad}, {Somerville},
  {Kravtsov}, {Klypin}, {Primack}  \& {Dekel}}{{Bullock}
  et~al.}{2001}]{2001MNRAS.321..559B}
{Bullock} J.~S.,  {Kolatt} T.~S.,  {Sigad} Y.,  {Somerville} R.~S.,  {Kravtsov}
  A.~V.,  {Klypin} A.~A.,  {Primack} J.~R.,   {Dekel} A.,  2001, \mn@doi
  [\mnras] {10.1046/j.1365-8711.2001.04068.x}, \href
  {http://adsabs.harvard.edu/abs/2001MNRAS.321..559B} {321, 559}

\bibitem[\protect\citeauthoryear{{Capaccioli} \& {Schipani}}{{Capaccioli} \&
  {Schipani}}{2011}]{2011Msngr.146....2C}
{Capaccioli} M.,  {Schipani} P.,  2011, The Messenger, \href
  {https://ui.adsabs.harvard.edu/abs/2011Msngr.146....2C} {146, 2}

\bibitem[\protect\citeauthoryear{{Cava} et~al.,}{{Cava}
  et~al.}{2009}]{2009A&A...495..707C}
{Cava} A.,  et~al., 2009, \mn@doi [\aap] {10.1051/0004-6361:200810997}, \href
  {https://ui.adsabs.harvard.edu/abs/2009A&A...495..707C} {495, 707}

\bibitem[\protect\citeauthoryear{{Child}, {Habib}, {Heitmann}, {Frontiere},
  {Finkel}, {Pope}  \& {Morozov}}{{Child} et~al.}{2018}]{2018ApJ...859...55C}
{Child} H.~L.,  {Habib} S.,  {Heitmann} K.,  {Frontiere} N.,  {Finkel} H.,
  {Pope} A.,   {Morozov} V.,  2018, \mn@doi [The Astrophysical Journal]
  {10.3847/1538-4357/aabf95}, \href
  {https://ui.adsabs.harvard.edu/abs/2018ApJ...859...55C} {859, 55}

\bibitem[\protect\citeauthoryear{{Chow-Mart{\'\i}nez}, {Andernach}, {Caretta}
  \& {Trejo-Alonso}}{{Chow-Mart{\'\i}nez} et~al.}{2014}]{2014MNRAS.445.4073C}
{Chow-Mart{\'\i}nez} M.,  {Andernach} H.,  {Caretta} C.~A.,   {Trejo-Alonso}
  J.~J.,  2014, \mn@doi [\mnras] {10.1093/mnras/stu1961}, \href
  {https://ui.adsabs.harvard.edu/abs/2014MNRAS.445.4073C} {445, 4073}

\bibitem[\protect\citeauthoryear{{Courtois}, {Tully}, {Hoffman},
  {Pomar{\`e}de}, {Graziani}  \& {Dupuy}}{{Courtois}
  et~al.}{2017}]{2017ApJ...847L...6C}
{Courtois} H.~M.,  {Tully} R.~B.,  {Hoffman} Y.,  {Pomar{\`e}de} D.,
  {Graziani} R.,   {Dupuy} A.,  2017, \mn@doi [\apjl]
  {10.3847/2041-8213/aa88b2}, \href
  {https://ui.adsabs.harvard.edu/abs/2017ApJ...847L...6C} {847, L6}

\bibitem[\protect\citeauthoryear{Diaferio \& Geller}{Diaferio \&
  Geller}{1997}]{Diaferio_1997}
Diaferio A.,  Geller M.~J.,  1997, \mn@doi [The Astrophysical Journal]
  {10.1086/304075}, 481, 633

\bibitem[\protect\citeauthoryear{{Dietrich}, {Werner}, {Clowe}, {Finoguenov},
  {Kitching}, {Miller}  \& {Simionescu}}{{Dietrich}
  et~al.}{2012}]{2012Natur.487..202D}
{Dietrich} J.~P.,  {Werner} N.,  {Clowe} D.,  {Finoguenov} A.,  {Kitching} T.,
  {Miller} L.,   {Simionescu} A.,  2012, \mn@doi [Nature]
  {10.1038/nature11224}, \href
  {http://adsabs.harvard.edu/abs/2012Natur.487..202D} {487, 202}

\bibitem[\protect\citeauthoryear{{Drinkwater}, {Parker}, {Proust}, {Slezak}  \&
  {Quintana}}{{Drinkwater} et~al.}{2004}]{2004PASA...21...89D}
{Drinkwater} M.~J.,  {Parker} Q.~A.,  {Proust} D.,  {Slezak} E.,   {Quintana}
  H.,  2004, \mn@doi [\pasa] {10.1071/AS03057}, \href
  {https://ui.adsabs.harvard.edu/abs/2004PASA...21...89D} {21, 89}

\bibitem[\protect\citeauthoryear{{Duffy}, {Schaye}, {Kay}  \& {Dalla
  Vecchia}}{{Duffy} et~al.}{2008}]{2008MNRAS.390L..64D}
{Duffy} A.~R.,  {Schaye} J.,  {Kay} S.~T.,   {Dalla Vecchia} C.,  2008, \mn@doi
  [MNRAS] {10.1111/j.1745-3933.2008.00537.x}, \href
  {http://adsabs.harvard.edu/abs/2008MNRAS.390L..64D} {390, L64}

\bibitem[\protect\citeauthoryear{{Einasto}, {H{\"u}tsi}, {Einasto}, {Saar},
  {Tucker}, {M{\"u}ller}, {Hein{\"a}m{\"a}ki}  \& {Allam}}{{Einasto}
  et~al.}{2003}]{2003A&A...405..425E}
{Einasto} J.,  {H{\"u}tsi} G.,  {Einasto} M.,  {Saar} E.,  {Tucker} D.~L.,
  {M{\"u}ller} V.,  {Hein{\"a}m{\"a}ki} P.,   {Allam} S.~S.,  2003, \mn@doi
  [\aap] {10.1051/0004-6361:20030419}, \href
  {http://adsabs.harvard.edu/abs/2003A%26A...405..425E} {405, 425}

\bibitem[\protect\citeauthoryear{{Einasto}, {Liivam{\"a}gi}, {Tago}, {Saar},
  {Tempel}, {Einasto}, {Mart{\'\i}nez}  \& {Hein{\"a}m{\"a}ki}}{{Einasto}
  et~al.}{2011}]{2011A&A...532A...5E}
{Einasto} M.,  {Liivam{\"a}gi} L.~J.,  {Tago} E.,  {Saar} E.,  {Tempel} E.,
  {Einasto} J.,  {Mart{\'\i}nez} V.~J.,   {Hein{\"a}m{\"a}ki} P.,  2011,
  \mn@doi [\aap] {10.1051/0004-6361/201116564}, \href
  {https://ui.adsabs.harvard.edu/abs/2011A&A...532A...5E} {532, A5}

\bibitem[\protect\citeauthoryear{{Einasto} et~al.,}{{Einasto}
  et~al.}{2016}]{2016A&A...595A..70E}
{Einasto} M.,  et~al., 2016, \mn@doi [\aap] {10.1051/0004-6361/201628567},
  \href {https://ui.adsabs.harvard.edu/abs/2016A&A...595A..70E} {595, A70}

\bibitem[\protect\citeauthoryear{{Einasto} et~al.,}{{Einasto}
  et~al.}{2018}]{2018A&A...620A.149E}
{Einasto} M.,  et~al., 2018, \mn@doi [\aap] {10.1051/0004-6361/201833711},
  \href {https://ui.adsabs.harvard.edu/abs/2018A&A...620A.149E} {620, A149}

\bibitem[\protect\citeauthoryear{{Einasto}, {Suhhonenko}, {Liivam{\"a}gi}  \&
  {Einasto}}{{Einasto} et~al.}{2019}]{2019A&A...623A..97E}
{Einasto} J.,  {Suhhonenko} I.,  {Liivam{\"a}gi} L.~J.,   {Einasto} M.,  2019,
  \mn@doi [\aap] {10.1051/0004-6361/201834450}, \href
  {https://ui.adsabs.harvard.edu/abs/2019A&A...623A..97E} {623, A97}

\bibitem[\protect\citeauthoryear{{Ettori}, {Fabian}  \& {White}}{{Ettori}
  et~al.}{1997}]{1997MNRAS.289..787E}
{Ettori} S.,  {Fabian} A.~C.,   {White} D.~A.,  1997, \mn@doi [\mnras]
  {10.1093/mnras/289.4.787}, \href
  {https://ui.adsabs.harvard.edu/abs/1997MNRAS.289..787E} {289, 787}

\bibitem[\protect\citeauthoryear{Ettori, Bardelli, de Grandi, Molendi, Zamorani
   \& Zucca}{Ettori et~al.}{2000}]{EBd00}
Ettori S.,  Bardelli S.,  de Grandi S.,  Molendi S.,  Zamorani G.,   Zucca E.,
  2000, \mn@doi [Monthly Notices of the Royal Astronomical Society]
  {10.1046/j.1365-8711.2000.03725.x}, 318, 239

\bibitem[\protect\citeauthoryear{Finoguenov, Henriksen, Briel, de Plaa  \&
  Kaastra}{Finoguenov et~al.}{2004}]{Finoguenov_2004}
Finoguenov A.,  Henriksen M.~J.,  Briel U.~G.,  de Plaa J.,   Kaastra J.~S.,
  2004, \mn@doi [The Astrophysical Journal] {10.1086/422246}, 611, 811

\bibitem[\protect\citeauthoryear{{Galametz} et~al.,}{{Galametz}
  et~al.}{2018}]{2018MNRAS.475.4148G}
{Galametz} A.,  et~al., 2018, \mn@doi [\mnras] {10.1093/mnras/sty095}, \href
  {https://ui.adsabs.harvard.edu/abs/2018MNRAS.475.4148G} {475, 4148}

\bibitem[\protect\citeauthoryear{{Giacintucci} et~al.,}{{Giacintucci}
  et~al.}{2005}]{2005A&A...440..867G}
{Giacintucci} S.,  et~al., 2005, \mn@doi [\aap] {10.1051/0004-6361:20053016},
  \href {https://ui.adsabs.harvard.edu/abs/2005A&A...440..867G} {440, 867}

\bibitem[\protect\citeauthoryear{{Giocoli}, {Meneghetti}, {Ettori}  \&
  {Moscardini}}{{Giocoli} et~al.}{2012}]{2012MNRAS.426.1558G}
{Giocoli} C.,  {Meneghetti} M.,  {Ettori} S.,   {Moscardini} L.,  2012, \mn@doi
  [\mnras] {10.1111/j.1365-2966.2012.21743.x}, \href
  {https://ui.adsabs.harvard.edu/abs/2012MNRAS.426.1558G} {426, 1558}

\bibitem[\protect\citeauthoryear{{Giocoli}, {Meneghetti}, {Metcalf}, {Ettori}
  \& {Moscardini}}{{Giocoli} et~al.}{2014}]{2014MNRAS.440.1899G}
{Giocoli} C.,  {Meneghetti} M.,  {Metcalf} R.~B.,  {Ettori} S.,   {Moscardini}
  L.,  2014, \mn@doi [\mnras] {10.1093/mnras/stu303}, \href
  {https://ui.adsabs.harvard.edu/abs/2014MNRAS.440.1899G} {440, 1899}

\bibitem[\protect\citeauthoryear{{Gonzalez}, {Zaritsky}  \&
  {Zabludoff}}{{Gonzalez} et~al.}{2007}]{2007ApJ...666..147G}
{Gonzalez} A.~H.,  {Zaritsky} D.,   {Zabludoff} A.~I.,  2007, \mn@doi [\apj]
  {10.1086/519729}, \href
  {https://ui.adsabs.harvard.edu/abs/2007ApJ...666..147G} {666, 147}

\bibitem[\protect\citeauthoryear{{Gonzalez}, {Sivanandam}, {Zabludoff}  \&
  {Zaritsky}}{{Gonzalez} et~al.}{2013}]{2013ApJ...778...14G}
{Gonzalez} A.~H.,  {Sivanandam} S.,  {Zabludoff} A.~I.,   {Zaritsky} D.,  2013,
  \mn@doi [\apj] {10.1088/0004-637X/778/1/14}, \href
  {https://ui.adsabs.harvard.edu/abs/2013ApJ...778...14G} {778, 14}

\bibitem[\protect\citeauthoryear{{Grado}, {Capaccioli}, {Limatola}  \&
  {Getman}}{{Grado} et~al.}{2012}]{2012MSAIS..19..362G}
{Grado} A.,  {Capaccioli} M.,  {Limatola} L.,   {Getman} F.,  2012, Memorie
  della Societa Astronomica Italiana Supplementi, \href
  {https://ui.adsabs.harvard.edu/abs/2012MSAIS..19..362G} {19, 362}

\bibitem[\protect\citeauthoryear{{Gray} et~al.,}{{Gray}
  et~al.}{2009}]{2009MNRAS.393.1275G}
{Gray} M.~E.,  et~al., 2009, \mn@doi [\mnras]
  {10.1111/j.1365-2966.2008.14259.x}, \href
  {https://ui.adsabs.harvard.edu/abs/2009MNRAS.393.1275G} {393, 1275}

\bibitem[\protect\citeauthoryear{{Haines} et~al.,}{{Haines}
  et~al.}{2018}]{2018MNRAS.481.1055H}
{Haines} C.~P.,  et~al., 2018, \mn@doi [\mnras] {10.1093/mnras/sty2338}, \href
  {https://ui.adsabs.harvard.edu/abs/2018MNRAS.481.1055H} {481, 1055}

\bibitem[\protect\citeauthoryear{{Heymans} et~al.,}{{Heymans}
  et~al.}{2008}]{2008MNRAS.385.1431H}
{Heymans} C.,  et~al., 2008, \mn@doi [MNRAS]
  {10.1111/j.1365-2966.2008.12919.x}, \href
  {http://adsabs.harvard.edu/abs/2008MNRAS.385.1431H} {385, 1431}

\bibitem[\protect\citeauthoryear{{Higuchi} \& {Inoue}}{{Higuchi} \&
  {Inoue}}{2019}]{2019MNRAS.488.5811H}
{Higuchi} Y.,  {Inoue} K.~T.,  2019, \mn@doi [\mnras] {10.1093/mnras/stz2150},
  \href {https://ui.adsabs.harvard.edu/abs/2019MNRAS.488.5811H} {488, 5811}

\bibitem[\protect\citeauthoryear{{Higuchi}, {Oguri}  \& {Shirasaki}}{{Higuchi}
  et~al.}{2014}]{2014MNRAS.441..745H}
{Higuchi} Y.,  {Oguri} M.,   {Shirasaki} M.,  2014, \mn@doi [MNRAS]
  {10.1093/mnras/stu583}, \href
  {http://adsabs.harvard.edu/abs/2014MNRAS.441..745H} {441, 745}

\bibitem[\protect\citeauthoryear{{Hinton}}{{Hinton}}{2016}]{Hinton2016}
{Hinton} S.~R.,  2016, \mn@doi [The Journal of Open Source Software]
  {10.21105/joss.00045}, \href
  {http://adsabs.harvard.edu/abs/2016JOSS....1...45H} {1, 00045}

\bibitem[\protect\citeauthoryear{{Ilbert} et~al.,}{{Ilbert}
  et~al.}{2013}]{2013A&A...556A..55I}
{Ilbert} O.,  et~al., 2013, \mn@doi [\aap] {10.1051/0004-6361/201321100}, \href
  {https://ui.adsabs.harvard.edu/abs/2013A&A...556A..55I} {556, A55}

\bibitem[\protect\citeauthoryear{{Jee}, {Hoekstra}, {Mahdavi}  \&
  {Babul}}{{Jee} et~al.}{2014}]{2014ApJ...783...78J}
{Jee} M.~J.,  {Hoekstra} H.,  {Mahdavi} A.,   {Babul} A.,  2014, \mn@doi [\apj]
  {10.1088/0004-637X/783/2/78}, \href
  {https://ui.adsabs.harvard.edu/abs/2014ApJ...783...78J} {783, 78}

\bibitem[\protect\citeauthoryear{{Jing}}{{Jing}}{2000}]{2000ApJ...535...30J}
{Jing} Y.~P.,  2000, \mn@doi [\apj] {10.1086/308809}, \href
  {http://adsabs.harvard.edu/abs/2000ApJ...535...30J} {535, 30}

\bibitem[\protect\citeauthoryear{{Jones} et~al.,}{{Jones}
  et~al.}{2009}]{2009MNRAS.399..683J}
{Jones} D.~H.,  et~al., 2009, \mn@doi [\mnras]
  {10.1111/j.1365-2966.2009.15338.x}, \href
  {https://ui.adsabs.harvard.edu/abs/2009MNRAS.399..683J} {399, 683}

\bibitem[\protect\citeauthoryear{{Kaiser} \& {Squires}}{{Kaiser} \&
  {Squires}}{1993}]{1993ApJ...404..441K}
{Kaiser} N.,  {Squires} G.,  1993, \mn@doi [\apj] {10.1086/172297}, \href
  {https://ui.adsabs.harvard.edu/abs/1993ApJ...404..441K} {404, 441}

\bibitem[\protect\citeauthoryear{{Kaiser}, {Squires}  \& {Broadhurst}}{{Kaiser}
  et~al.}{1995}]{1995ApJ...449..460K}
{Kaiser} N.,  {Squires} G.,   {Broadhurst} T.,  1995, \mn@doi [\apj]
  {10.1086/176071}, \href
  {https://ui.adsabs.harvard.edu/abs/1995ApJ...449..460K} {449, 460}

\bibitem[\protect\citeauthoryear{{Kaldare}, {Colless}, {Raychaudhury}  \&
  {Peterson}}{{Kaldare} et~al.}{2003}]{2003MNRAS.339..652K}
{Kaldare} R.,  {Colless} M.,  {Raychaudhury} S.,   {Peterson} B.~A.,  2003,
  \mn@doi [\mnras] {10.1046/j.1365-8711.2003.05695.x}, \href
  {https://ui.adsabs.harvard.edu/abs/2003MNRAS.339..652K} {339, 652}

\bibitem[\protect\citeauthoryear{{Kocevski}, {Ebeling}  \& {Mullis}}{{Kocevski}
  et~al.}{2004}]{2004cgpc.sympE..26K}
{Kocevski} D.~D.,  {Ebeling} H.,   {Mullis} C.~R.,  2004, in {Mulchaey} J.~S.,
  {Dressler} A.,   {Oemler} A.,  eds, Clusters of Galaxies: Probes of
  Cosmological Structure and Galaxy Evolution. p.~26 (\mn@eprint {arXiv}
  {astro-ph/0304453})

\bibitem[\protect\citeauthoryear{{Kron}}{{Kron}}{1980}]{1980ApJS...43..305K}
{Kron} R.~G.,  1980, \mn@doi [\apjs] {10.1086/190669}, \href
  {https://ui.adsabs.harvard.edu/abs/1980ApJS...43..305K} {43, 305}

\bibitem[\protect\citeauthoryear{{Kuijken}}{{Kuijken}}{2011}]{2011Msngr.146....8K}
{Kuijken} K.,  2011, The Messenger, \href
  {https://ui.adsabs.harvard.edu/abs/2011Msngr.146....8K} {146, 8}

\bibitem[\protect\citeauthoryear{{Kull} \& {B{\"o}hringer}}{{Kull} \&
  {B{\"o}hringer}}{1999}]{1999A&A...341...23K}
{Kull} A.,  {B{\"o}hringer} H.,  1999, \aap, \href
  {https://ui.adsabs.harvard.edu/abs/1999A&A...341...23K} {341, 23}

\bibitem[\protect\citeauthoryear{{Lubin}, {Gal}, {Lemaux}, {Kocevski}  \&
  {Squires}}{{Lubin} et~al.}{2009}]{2009AJ....137.4867L}
{Lubin} L.~M.,  {Gal} R.~R.,  {Lemaux} B.~C.,  {Kocevski} D.~D.,   {Squires}
  G.~K.,  2009, \mn@doi [\aj] {10.1088/0004-6256/137/6/4867}, \href
  {https://ui.adsabs.harvard.edu/abs/2009AJ....137.4867L} {137, 4867}

\bibitem[\protect\citeauthoryear{{Mahajan}, {Singh}  \& {Shobhana}}{{Mahajan}
  et~al.}{2018}]{2018MNRAS.478.4336M}
{Mahajan} S.,  {Singh} A.,   {Shobhana} D.,  2018, \mn@doi [\mnras]
  {10.1093/mnras/sty1370}, \href
  {https://ui.adsabs.harvard.edu/abs/2018MNRAS.478.4336M} {478, 4336}

\bibitem[\protect\citeauthoryear{{Maturi} \& {Merten}}{{Maturi} \&
  {Merten}}{2013}]{2013A&A...559A.112M}
{Maturi} M.,  {Merten} J.,  2013, \mn@doi [\aap] {10.1051/0004-6361/201322007},
  \href {https://ui.adsabs.harvard.edu/abs/2013A&A...559A.112M} {559, A112}

\bibitem[\protect\citeauthoryear{{Medezinski}, {Broadhurst}, {Umetsu}, {Oguri},
  {Rephaeli}  \& {Ben{\'{\i}}tez}}{{Medezinski}
  et~al.}{2010}]{2010MNRAS.405..257M}
{Medezinski} E.,  {Broadhurst} T.,  {Umetsu} K.,  {Oguri} M.,  {Rephaeli} Y.,
  {Ben{\'{\i}}tez} N.,  2010, \mn@doi [\mnras]
  {10.1111/j.1365-2966.2010.16491.x}, \href
  {http://adsabs.harvard.edu/abs/2010MNRAS.405..257M} {405, 257}

\bibitem[\protect\citeauthoryear{{Mei} et~al.,}{{Mei}
  et~al.}{2012}]{2012ApJ...754..141M}
{Mei} S.,  et~al., 2012, \mn@doi [\apj] {10.1088/0004-637X/754/2/141}, \href
  {https://ui.adsabs.harvard.edu/abs/2012ApJ...754..141M} {754, 141}

\bibitem[\protect\citeauthoryear{{Meneghetti}, {Rasia}, {Merten}, {Bellagamba},
  {Ettori}, {Mazzotta}, {Dolag}  \& {Marri}}{{Meneghetti}
  et~al.}{2010}]{2010A&A...514A..93M}
{Meneghetti} M.,  {Rasia} E.,  {Merten} J.,  {Bellagamba} F.,  {Ettori} S.,
  {Mazzotta} P.,  {Dolag} K.,   {Marri} S.,  2010, \mn@doi [\aap]
  {10.1051/0004-6361/200913222}, \href
  {https://ui.adsabs.harvard.edu/abs/2010A&A...514A..93M} {514, A93}

\bibitem[\protect\citeauthoryear{{Mercurio} et~al.,}{{Mercurio}
  et~al.}{2015}]{2015MNRAS.453.3685M}
{Mercurio} A.,  et~al., 2015, \mn@doi [\mnras] {10.1093/mnras/stv1905}, \href
  {https://ui.adsabs.harvard.edu/abs/2015MNRAS.453.3685M} {453, 3685}

\bibitem[\protect\citeauthoryear{{Merluzzi} et~al.,}{{Merluzzi}
  et~al.}{2015}]{2015MNRAS.446..803M}
{Merluzzi} P.,  et~al., 2015, \mn@doi [\mnras] {10.1093/mnras/stu2085}, \href
  {http://adsabs.harvard.edu/abs/2015MNRAS.446..803M} {446, 803}

\bibitem[\protect\citeauthoryear{{Merluzzi}, {Busarello}, {Dopita}, {Haines},
  {Steinhauser}, {Bourdin}  \& {Mazzotta}}{{Merluzzi}
  et~al.}{2016}]{2016MNRAS.460.3345M}
{Merluzzi} P.,  {Busarello} G.,  {Dopita} M.~A.,  {Haines} C.~P.,
  {Steinhauser} D.,  {Bourdin} H.,   {Mazzotta} P.,  2016, \mn@doi [\mnras]
  {10.1093/mnras/stw1198}, \href
  {https://ui.adsabs.harvard.edu/abs/2016MNRAS.460.3345M} {460, 3345}

\bibitem[\protect\citeauthoryear{{Miller}}{{Miller}}{2005}]{2005AJ....130.2541M}
{Miller} N.~A.,  2005, \mn@doi [\aj] {10.1086/497165}, \href
  {https://ui.adsabs.harvard.edu/abs/2005AJ....130.2541M} {130, 2541}

\bibitem[\protect\citeauthoryear{{Mu{\~n}oz} \& {Loeb}}{{Mu{\~n}oz} \&
  {Loeb}}{2008}]{2008MNRAS.391.1341M}
{Mu{\~n}oz} J.~A.,  {Loeb} A.,  2008, \mn@doi [\mnras]
  {10.1111/j.1365-2966.2008.13973.x}, \href
  {https://ui.adsabs.harvard.edu/abs/2008MNRAS.391.1341M} {391, 1341}

\bibitem[\protect\citeauthoryear{{Navarro}, {Frenk}  \& {White}}{{Navarro}
  et~al.}{1997}]{1997ApJ...490..493N}
{Navarro} J.~F.,  {Frenk} C.~S.,   {White} S.~D.~M.,  1997, \mn@doi [ApJ]
  {10.1086/304888}, \href {http://adsabs.harvard.edu/abs/1997ApJ...490..493N}
  {490, 493}

\bibitem[\protect\citeauthoryear{{Neto} et~al.,}{{Neto}
  et~al.}{2007}]{2007MNRAS.381.1450N}
{Neto} A.~F.,  et~al., 2007, \mn@doi [Monthly Notices of the Royal Astronomical
  Society] {10.1111/j.1365-2966.2007.12381.x}, \href
  {https://ui.adsabs.harvard.edu/abs/2007MNRAS.381.1450N} {381, 1450}

\bibitem[\protect\citeauthoryear{{Oguri}, {Takahashi}, {Ichiki}  \&
  {Ohno}}{{Oguri} et~al.}{2004}]{2004astro.ph.10145O}
{Oguri} M.,  {Takahashi} K.,  {Ichiki} K.,   {Ohno} H.,  2004, arXiv
  Astrophysics e-prints, \href
  {http://adsabs.harvard.edu/abs/2004astro.ph.10145O} {}

\bibitem[\protect\citeauthoryear{{Oguri}, {Takada}, {Umetsu}  \&
  {Broadhurst}}{{Oguri} et~al.}{2005}]{2005ApJ...632..841O}
{Oguri} M.,  {Takada} M.,  {Umetsu} K.,   {Broadhurst} T.,  2005, \mn@doi
  [\apj] {10.1086/452629}, \href
  {https://ui.adsabs.harvard.edu/abs/2005ApJ...632..841O} {632, 841}

\bibitem[\protect\citeauthoryear{{Oguri}, {Takada}, {Okabe}  \&
  {Smith}}{{Oguri} et~al.}{2010}]{2010MNRAS.405.2215O}
{Oguri} M.,  {Takada} M.,  {Okabe} N.,   {Smith} G.~P.,  2010, \mn@doi [\mnras]
  {10.1111/j.1365-2966.2010.16622.x}, \href
  {http://adsabs.harvard.edu/abs/2010MNRAS.405.2215O} {405, 2215}

\bibitem[\protect\citeauthoryear{{Oguri}, {Bayliss}, {Dahle}, {Sharon},
  {Gladders}, {Natarajan}, {Hennawi}  \& {Koester}}{{Oguri}
  et~al.}{2012}]{2012MNRAS.420.3213O}
{Oguri} M.,  {Bayliss} M.~B.,  {Dahle} H.,  {Sharon} K.,  {Gladders} M.~D.,
  {Natarajan} P.,  {Hennawi} J.~F.,   {Koester} B.~P.,  2012, \mn@doi [MNRAS]
  {10.1111/j.1365-2966.2011.20248.x}, \href
  {http://adsabs.harvard.edu/abs/2012MNRAS.420.3213O} {420, 3213}

\bibitem[\protect\citeauthoryear{{Okabe} \& {Umetsu}}{{Okabe} \&
  {Umetsu}}{2008}]{2008PASJ...60..345O}
{Okabe} N.,  {Umetsu} K.,  2008, \mn@doi [\pasj] {10.1093/pasj/60.2.345}, \href
  {https://ui.adsabs.harvard.edu/abs/2008PASJ...60..345O} {60, 345}

\bibitem[\protect\citeauthoryear{{Okabe}, {Takada}, {Umetsu}, {Futamase}  \&
  {Smith}}{{Okabe} et~al.}{2010}]{2010PASJ...62..811O}
{Okabe} N.,  {Takada} M.,  {Umetsu} K.,  {Futamase} T.,   {Smith} G.~P.,  2010,
  PASJ, \href {http://adsabs.harvard.edu/abs/2010PASJ...62..811O} {62, 811}

\bibitem[\protect\citeauthoryear{{Okabe}, {Smith}, {Umetsu}, {Takada}  \&
  {Futamase}}{{Okabe} et~al.}{2013}]{2013ApJ...769L..35O}
{Okabe} N.,  {Smith} G.~P.,  {Umetsu} K.,  {Takada} M.,   {Futamase} T.,  2013,
  \mn@doi [\apjl] {10.1088/2041-8205/769/2/L35}, \href
  {https://ui.adsabs.harvard.edu/abs/2013ApJ...769L..35O} {769, L35}

\bibitem[\protect\citeauthoryear{{Okabe}, {Futamase}, {Kajisawa}  \&
  {Kuroshima}}{{Okabe} et~al.}{2014}]{2014ApJ...784...90O}
{Okabe} N.,  {Futamase} T.,  {Kajisawa} M.,   {Kuroshima} R.,  2014, \mn@doi
  [\apj] {10.1088/0004-637X/784/2/90}, \href
  {https://ui.adsabs.harvard.edu/abs/2014ApJ...784...90O} {784, 90}

\bibitem[\protect\citeauthoryear{{Okabe} et~al.,}{{Okabe}
  et~al.}{2016}]{2016MNRAS.456.4475O}
{Okabe} N.,  et~al., 2016, \mn@doi [\mnras] {10.1093/mnras/stv2916}, \href
  {https://ui.adsabs.harvard.edu/abs/2016MNRAS.456.4475O} {456, 4475}

\bibitem[\protect\citeauthoryear{{Okabe} et~al.,}{{Okabe}
  et~al.}{2019}]{2019PASJ...71...79O}
{Okabe} N.,  et~al., 2019, \mn@doi [Publications of the Astronomical Society of
  Japan] {10.1093/pasj/psz059}, \href
  {https://ui.adsabs.harvard.edu/abs/2019PASJ...71...79O} {71, 79}

\bibitem[\protect\citeauthoryear{{Planck Collaboration} et~al.,}{{Planck
  Collaboration} et~al.}{2014}]{2014A&A...571A..31P}
{Planck Collaboration} et~al., 2014, \mn@doi [\aap]
  {10.1051/0004-6361/201423743}, \href
  {https://ui.adsabs.harvard.edu/abs/2014A&A...571A..31P} {571, A31}

\bibitem[\protect\citeauthoryear{{Pratt}, {Arnaud}, {Biviano}, {Eckert},
  {Ettori}, {Nagai}, {Okabe}  \& {Reiprich}}{{Pratt}
  et~al.}{2019}]{2019SSRv..215...25P}
{Pratt} G.~W.,  {Arnaud} M.,  {Biviano} A.,  {Eckert} D.,  {Ettori} S.,
  {Nagai} D.,  {Okabe} N.,   {Reiprich} T.~H.,  2019, \mn@doi [\ssr]
  {10.1007/s11214-019-0591-0}, \href
  {https://ui.adsabs.harvard.edu/abs/2019SSRv..215...25P} {215, 25}

\bibitem[\protect\citeauthoryear{{Proust} et~al.,}{{Proust}
  et~al.}{2006}]{2006A&A...447..133P}
{Proust} D.,  et~al., 2006, \mn@doi [\aap] {10.1051/0004-6361:20052838}, \href
  {https://ui.adsabs.harvard.edu/abs/2006A&A...447..133P} {447, 133}

\bibitem[\protect\citeauthoryear{{Quintana}, {Ramirez}, {Melnick},
  {Raychaudhury}  \& {Slezak}}{{Quintana} et~al.}{1995}]{1995AJ....110..463Q}
{Quintana} H.,  {Ramirez} A.,  {Melnick} J.,  {Raychaudhury} S.,   {Slezak} E.,
   1995, \mn@doi [\aj] {10.1086/117535}, \href
  {https://ui.adsabs.harvard.edu/abs/1995AJ....110..463Q} {110, 463}

\bibitem[\protect\citeauthoryear{{Quintana}, {Melnick}, {Proust}  \&
  {Infante}}{{Quintana} et~al.}{1997}]{1997A&AS..125..247Q}
{Quintana} H.,  {Melnick} J.,  {Proust} D.,   {Infante} L.,  1997, \mn@doi
  [\aaps] {10.1051/aas:1997218}, \href
  {https://ui.adsabs.harvard.edu/abs/1997A&AS..125..247Q} {125, 247}

\bibitem[\protect\citeauthoryear{{Ragone}, {Muriel}, {Proust}, {Reisenegger}
  \& {Quintana}}{{Ragone} et~al.}{2006}]{RMP06}
{Ragone} C.~J.,  {Muriel} H.,  {Proust} D.,  {Reisenegger} A.,   {Quintana} H.,
   2006, \mn@doi [A\&A] {10.1051/0004-6361:20053623}, 445, 819

\bibitem[\protect\citeauthoryear{{Raychaudhury}}{{Raychaudhury}}{1989}]{1989Natur.342..251R}
{Raychaudhury} S.,  1989, \mn@doi [\nat] {10.1038/342251a0}, \href
  {https://ui.adsabs.harvard.edu/abs/1989Natur.342..251R} {342, 251}

\bibitem[\protect\citeauthoryear{{Reisenegger}, {Quintana}, {Carrasco}  \&
  {Maze}}{{Reisenegger} et~al.}{2000}]{2000AJ....120..523R}
{Reisenegger} A.,  {Quintana} H.,  {Carrasco} E.~R.,   {Maze} J.,  2000,
  \mn@doi [\aj] {10.1086/301477}, \href
  {https://ui.adsabs.harvard.edu/abs/2000AJ....120..523R} {120, 523}

\bibitem[\protect\citeauthoryear{{Rossetti}, {Ghizzardi}, {Molendi}  \&
  {Finoguenov}}{{Rossetti} et~al.}{2007b}]{2007A&A...463..839R}
{Rossetti} M.,  {Ghizzardi} S.,  {Molendi} S.,   {Finoguenov} A.,  2007b,
  \mn@doi [\aap] {10.1051/0004-6361:20054621}, \href
  {https://ui.adsabs.harvard.edu/abs/2007A&A...463..839R} {463, 839}

\bibitem[\protect\citeauthoryear{{Rossetti}, {Ghizzardi}, {Molendi}  \&
  {Finoguenov}}{{Rossetti} et~al.}{2007a}]{RGM07}
{Rossetti} M.,  {Ghizzardi} S.,  {Molendi} S.,   {Finoguenov} A.,  2007a,
  \mn@doi [A\&A] {10.1051/0004-6361:20054621}, 463, 839

\bibitem[\protect\citeauthoryear{{Schlafly} \& {Finkbeiner}}{{Schlafly} \&
  {Finkbeiner}}{2011}]{2011ApJ...737..103S}
{Schlafly} E.~F.,  {Finkbeiner} D.~P.,  2011, \mn@doi [\apj]
  {10.1088/0004-637X/737/2/103}, \href
  {https://ui.adsabs.harvard.edu/abs/2011ApJ...737..103S} {737, 103}

\bibitem[\protect\citeauthoryear{{Sereno} \& {Ettori}}{{Sereno} \&
  {Ettori}}{2015}]{2015MNRAS.450.3633S}
{Sereno} M.,  {Ettori} S.,  2015, \mn@doi [\mnras] {10.1093/mnras/stv810},
  \href {https://ui.adsabs.harvard.edu/abs/2015MNRAS.450.3633S} {450, 3633}

\bibitem[\protect\citeauthoryear{{Sereno} \& {Umetsu}}{{Sereno} \&
  {Umetsu}}{2011}]{2011MNRAS.416.3187S}
{Sereno} M.,  {Umetsu} K.,  2011, \mn@doi [\mnras]
  {10.1111/j.1365-2966.2011.19274.x}, \href
  {https://ui.adsabs.harvard.edu/abs/2011MNRAS.416.3187S} {416, 3187}

\bibitem[\protect\citeauthoryear{{Takizawa}, {Nagino}  \&
  {Matsushita}}{{Takizawa} et~al.}{2010}]{2010PASJ...62..951T}
{Takizawa} M.,  {Nagino} R.,   {Matsushita} K.,  2010, \mn@doi [\pasj]
  {10.1093/pasj/62.4.951}, \href
  {https://ui.adsabs.harvard.edu/abs/2010PASJ...62..951T} {62, 951}

\bibitem[\protect\citeauthoryear{{Tully}}{{Tully}}{2005}]{2005ApJ...618..214T}
{Tully} R.~B.,  2005, \mn@doi [\apj] {10.1086/425852}, \href
  {https://ui.adsabs.harvard.edu/abs/2005ApJ...618..214T} {618, 214}

\bibitem[\protect\citeauthoryear{{Umetsu}, {Medezinski}, {Broadhurst},
  {Zitrin}, {Okabe}, {Hsieh}  \& {Molnar}}{{Umetsu}
  et~al.}{2010}]{2010ApJ...714.1470U}
{Umetsu} K.,  {Medezinski} E.,  {Broadhurst} T.,  {Zitrin} A.,  {Okabe} N.,
  {Hsieh} B.-C.,   {Molnar} S.~M.,  2010, \mn@doi [\apj]
  {10.1088/0004-637X/714/2/1470}, \href
  {https://ui.adsabs.harvard.edu/abs/2010ApJ...714.1470U} {714, 1470}

\bibitem[\protect\citeauthoryear{{Umetsu}, {Broadhurst}, {Zitrin},
  {Medezinski}, {Coe}  \& {Postman}}{{Umetsu}
  et~al.}{2011}]{2011ApJ...738...41U}
{Umetsu} K.,  {Broadhurst} T.,  {Zitrin} A.,  {Medezinski} E.,  {Coe} D.,
  {Postman} M.,  2011, \mn@doi [\apj] {10.1088/0004-637X/738/1/41}, \href
  {http://adsabs.harvard.edu/abs/2011ApJ...738...41U} {738, 41}

\bibitem[\protect\citeauthoryear{{Umetsu} et~al.,}{{Umetsu}
  et~al.}{2014}]{2014ApJ...795..163U}
{Umetsu} K.,  et~al., 2014, \mn@doi [\apj] {10.1088/0004-637X/795/2/163}, \href
  {http://adsabs.harvard.edu/abs/2014ApJ...795..163U} {795, 163}

\bibitem[\protect\citeauthoryear{{Umetsu}, {Zitrin}, {Gruen}, {Merten},
  {Donahue}  \& {Postman}}{{Umetsu} et~al.}{2016}]{2016ApJ...821..116U}
{Umetsu} K.,  {Zitrin} A.,  {Gruen} D.,  {Merten} J.,  {Donahue} M.,
  {Postman} M.,  2016, \mn@doi [\apj] {10.3847/0004-637X/821/2/116}, \href
  {http://adsabs.harvard.edu/abs/2016ApJ...821..116U} {821, 116}

\bibitem[\protect\citeauthoryear{{Venturi}, {Bardelli}, {Dallacasa},
  {Brunetti}, {Giacintucci}, {Hunstead}  \& {Morganti}}{{Venturi}
  et~al.}{2003}]{2003A&A...402..913V}
{Venturi} T.,  {Bardelli} S.,  {Dallacasa} D.,  {Brunetti} G.,  {Giacintucci}
  S.,  {Hunstead} R.~W.,   {Morganti} R.,  2003, \mn@doi [\aap]
  {10.1051/0004-6361:20030345}, \href
  {https://ui.adsabs.harvard.edu/abs/2003A&A...402..913V} {402, 913}

\bibitem[\protect\citeauthoryear{{Wright} \& {Brainerd}}{{Wright} \&
  {Brainerd}}{2000}]{2000ApJ...534...34W}
{Wright} C.~O.,  {Brainerd} T.~G.,  2000, \mn@doi [ApJ] {10.1086/308744}, \href
  {http://adsabs.harvard.edu/abs/2000ApJ...534...34W} {534, 34}

\bibitem[\protect\citeauthoryear{{Yaryura}, {Baugh}  \& {Angulo}}{{Yaryura}
  et~al.}{2011}]{2011MNRAS.413.1311Y}
{Yaryura} C.~Y.,  {Baugh} C.~M.,   {Angulo} R.~E.,  2011, \mn@doi [\mnras]
  {10.1111/j.1365-2966.2011.18233.x}, \href
  {https://ui.adsabs.harvard.edu/abs/2011MNRAS.413.1311Y} {413, 1311}

\bibitem[\protect\citeauthoryear{{York} et~al.,}{{York}
  et~al.}{2000}]{2000AJ....120.1579Y}
{York} D.~G.,  et~al., 2000, \mn@doi [AJ] {10.1086/301513}, \href
  {http://adsabs.harvard.edu/abs/2000AJ....120.1579Y} {120, 1579}

\bibitem[\protect\citeauthoryear{{Zhao}, {Jing}, {Mo}  \& {B{\"o}rner}}{{Zhao}
  et~al.}{2009}]{2009ApJ...707..354Z}
{Zhao} D.~H.,  {Jing} Y.~P.,  {Mo} H.~J.,   {B{\"o}rner} G.,  2009, \mn@doi
  [\apj] {10.1088/0004-637X/707/1/354}, \href
  {http://adsabs.harvard.edu/abs/2009ApJ...707..354Z} {707, 354}

\bibitem[\protect\citeauthoryear{{de Filippis}, {Schindler}  \& {Erben}}{{de
  Filippis} et~al.}{2005}]{2005A&A...444..387D}
{de Filippis} E.,  {Schindler} S.,   {Erben} T.,  2005, \mn@doi [\aap]
  {10.1051/0004-6361:20053675}, \href
  {https://ui.adsabs.harvard.edu/abs/2005A&A...444..387D} {444, 387}

\makeatother
\end{thebibliography}




\bsp	
\label{lastpage}
\end{document}